\documentclass[12pt,preprint]{aastex}

\newcommand{\nc}{\newcommand}

\nc{\lsun}{\ensuremath{\mathrm{L}_\odot}}
\nc{\msun}{\ensuremath{\mathrm{M}_\odot}}
\nc{\kms}{\mbox{km~s$^{-1}$}}
\def\arcsec{{$^{\prime\prime}$}}

\def\psec{$^s\mskip-7.6mu.\,$}
\def\Msun{\,{\rm M$_{\odot}$}}
\def\Lsun{\,{\rm L$_{\odot}$}}
\nc{\tdust}{\mbox{$T_{\rm d}$}}
\nc{\tstar}{\mbox{$T_{\ast}$}}
\nc{\mstar}{\mbox{$M_{\ast}$}}
\nc{\lstar}{\mbox{$L_{\ast}$}}
\nc{\tex}{\mbox{$T_{\rm ex}$}}
\nc{\twCO}{$^{12}$CO}
\nc{\av}{$A_{\rm v}$}
\nc{\halpha}{H$\alpha$}

\slugcomment{Accepted  8 October 2009}

\shorttitle{YSOs in  Ori I-2}
\shortauthors{Mookerjea \& Sandell}

\begin{document}


\title{Star formation in  the Cometary Globule Ori I-2}


\author{Bhaswati Mookerjea}
\affil{Department of Astronomy \& Astrophysics, Tata Institute of
Fundamental Research, Homi Bhabha Road, Mumbai 400005, India}
\email{bhaswati@tifr.res.in}

\and

\author{G\"oran Sandell}
\affil{SOFIA-USRA, NASA Ames Research Center, MS 211-3,
Moffett Field, CA 94035, USA}
\email{Goran.H.Sandell@nasa.gov}

\begin{abstract}

We investigate the young stellar population in and near the cometary
globule Ori\,I-2. The analysis is based on deep Nordic Optical Telescope
$R$-band and H$\alpha$ images,  JCMT SCUBA 450 and 850~\micron\ images
combined with near-infrared 2MASS photometry and mid-infrared archival
{\em Spitzer} images obtained with the IRAC (3.6, 4.5, 5.8 and
8~\micron), and MIPS (24 and 70~\micron) instruments. We identify a
total of 125 sources within the 5\arcmin$\times$5\arcmin\ region imaged
by IRAC. Of these sources 87 are detected in the $R$-band image and 51
are detected in the 2MASS survey. The detailed physical properties of
the sources are explored using a combination of near/mid-infrared
color-color diagrams, greybody fitting of SEDs and an online SED fitting
tool that uses a library of 2D radiation transfer based accretion models
of young stellar objects with disks. Ori\,I-2 shows clear evidence of
triggered star formation with four young low luminosity pre-main
sequence stars embedded in the globule. At least two, possibly as many
as four, additional low-mass PMS objects, were discovered in the field
which are probably part of the young $\sigma$ Orionis cluster.  Among
the PMS stars which have formed in the globule, MIR-54 is a young,
deeply embedded Class 0/I object, MIR-51 and 52 are young Class II
sources, while MIR-89 is a more evolved, heavily extincted Class II
object with its apparent colors mimicking a Class 0/I object.  The Class
0/I  object MIR-54 coincides with a previously known IRAS source and is
a strong sub-millimeter source. It is most likely the source for the
molecular outflow and the large parsec scale Herbig-Haro flow. However
the nearby Class II source, MIR-52, which is strong a H$\alpha$ emission
line star, also appears to drive an outflow approximately aligned with
the outflow from MIR-54, and because of the proximity of the two
outflows, either star could contribute.  MIR-89 appears to excite a low
excitation HH object, HH~992, discovered for the first time in this
study.

\end{abstract}


\section{Introduction}

Cometary globules are believed to be molecular cloud condensations,
which are so dense that they are not disrupted when an \ion{H}{2}
region expands into  the molecular cloud(s) surrounding it.  Such
globules are always bright rimmed and often have a cometary or tear
shaped form, because the stellar wind and ionizing radiation from the
central O stars ablate away the low density gas on the side facing the
O stars and sweep away dust towards the tail. Cometary globules are
thought to form from ``elephant trunks''  \citep{Herbig74,Reipurth84},
although for compact cores surrounded by low density gas this phase
can be quite short. The Rosette nebula is a good example of an
\ion{H}{2} region, which shows an abundance of both cometary globules
and elephant trunk structures \citep{Herbig74}. Although elephant
trunks and cometary globules have been known for a long time, the fact that
these sites may be forming stars was realized only in the last few
decades.  The first clear confirmation that stars form in cometary
globules was a result of the identification of Bernes~135 in the
cometary globule CG\,1, one of the large cometary globules in the Gum
Nebula, as a pre-main-sequence (PMS) star \citep{Reipurth83}.  The discovery of molecular outflows associated
with cold IRAS sources without optical counterparts in three
bright-rimmed globules \citep{sugitani1989}, definitely confirmed that
stars form in elephant trunks and cometary globules. Probably the most
well-known example of star forming in the tip of elephant trunks is seen in
the Hubble poster,  ``the pillars of creation'',  in the Eagle nebula (M\,16) \citep{Hester96}.

In this paper we examine star formation  in a more nearby 
cometary globule, Ori I-2, located in the large \ion{H}{2} region IC\,434.
This is one of the three globules, in which \citet{sugitani1989}
discovered a molecular outflow.  By using {\it Spitzer} IRAC and MIPS
images, 2MASS data, deep R and H$\alpha$ images from the Nordic
Optical Telescope (NOT), and SCUBA sub-millimeter imaging we can
identify all young stars in and near the globule and provide more
accurate information on their physical characteristics.

\section{Overview of previous work on Ori\,I-2}

\citet{Bok71} were the first to notice this globule in the large
\ion{H}{2} region IC\,434 and named it Ori\,I-2. The \ion{H}{2} region
IC\,434 is illuminated by the Trapezium like system $\sigma$~Ori with at
least five early type stars \citep{caballero2008}. The two hottest
members, A and B (O9.5V and B0.5V) form a binary system, which allows
for an accurate distance determination.  \citet{caballero2008} derived a
distance of 334 pc assuming that they form a binary, although he also
considered the possibility that it is a hierarchical triple system, in
which case the distance would be $\sim$ 385 pc. \citet{walter2008}, in their review of distance
estimates to the $\sigma$ Orionis cluster, quote a slightly larger distance, 420 $\pm$
30 pc. The morphology of Ori\,I-2, with the bright rim facing  $\sigma$~Ori  and the tail aligned in
the direction of the star suggests that the globule is at the same
distance or slightly in front of  $\sigma$~Ori. In this paper we adopt
the distance to Ori\,I-2 as 380 pc, favoring the distance estimate by \citeauthor{caballero2008}. 

 
\citet{Martin78} included Ori\,I-2 in their  molecular line survey of
small Bok globules and found  it to be the densest,  the most
compact, and the warmest, 18~K, of all the globules in their sample. \citet{sugitani1989}
discovered a bipolar molecular outflow in the globule. They found that
the outflow had a linear extent of 0.34 pc,  a dynamical age of 5
$\times$ 10$^4$ yr, and was centered on a low luminosity (L = 11
\Lsun{}) cold IRAS source, IRAS~$05355-0146$, without any optical
counterpart.  \citet{cernicharo1992} did a more detailed study of
Ori\,I-2 in CO, $^{13}$CO and C$^{18}$O J = $1 \to 0$,  HCN J = $1 \to
0$, CS J = $2 \to 1$ and CS  J = $3 \to 2$ with higher spatial
resolution and sensitivity than \citet{Martin78} and
\citet{sugitani1989}. They confirmed that the molecular outflow is
bipolar and  centered on the IRAS source, IRAS~$05355-0146$, but found
it to be more compact (0.22~pc) than what \citet{sugitani1989}
estimated from their low-spatial resolution observations. By examining
the Palomar sky survey prints and doing CCD imaging in R, I, and
narrowband H$\alpha$ and [\ion{S}{2}], they showed that the bright rim
is a diffuse \ion{H}{2} region and not a reflection nebula. They
identified a faint star, star A, as a possible counterpart to the IRAS
source. They found that the gas temperature of the shock heated gas in
the bright rim is $\sim$ 25 K, while the gas temperature in the opaque
(A$_V$ $\sim$ 20$^m$) core of the globule is  $\sim$ 12 K, similar to
that of normal isolated globules. They determined the total mass of
the globule to be $\sim$ 8 \Msun, with the opaque core having a mass
of $\sim$ 2.3 \Msun\ (corrected to a distance of 380 pc). Ori\,I-2 was
included in the catalogue of remnant molecular clouds in the Ori OB\,1
association by \citet{ogura1998} as cloud number 40A. They also
identified three fainter cometary globules  ``behind''  the tail of
Ori\,I-2 (40\,A), which they called 40\,B, C, and D.
\citet{mader1999}, who did a photographic and CCD imaging survey of
the L~1630 and L~1641 molecular cloud regions in Orion, identified a
chain of HH objects, HH\,289, to the east of Ori I-2 and a tube like
feature (cavity) protruding out on the western side of the globule,
both apparently excited by the embedded IRAS source.  Their
observations show that the  IRAS source drives a giant  outflow with a
projected size of $\sim 1$~pc to the east. A good overview of Ori\,I-2
and other globules in the vicinity of $\sigma$ Orionis can be found in
\citet{alcala2008}.

\section{Observations}

\subsection{{\em Spitzer} IRAC \& MIPS  observations}

We have extracted IRAC and MIPS (24 \& 70~\micron) observations from
the Spitzer Space Observatory archive (Program ID 30050: Star
Formation in Bright Rimmed Clouds by Fazio et al.).  The IRAC data
were taken in the High Dynamic Readout (HDR) mode  using a single AOR
(Astronomical Observation Request) with a five-point dither pattern.
We have processed both the short (0.6~sec) and the long (12~sec)
integration Basic Calibrated Data (BCD) frames in each channel using
the Artifact mitigation software developed by Sean Carey and created
mosaics using MOPEX.  These IRAC observations go far deeper than any
previous observations of Ori\,I-2.  In the short wavelength bands we
can detect point-sources down to 30 -- 40~$\mu$Jy and are slightly
less sensitive in the long wavelength bands, but in regions without
nebulosity we can still detect point-sources down to $\sim$  60
$\mu$Jy.

We have also created mosaics of the  MIPS 24 and 70~\micron\   BCDs
using MOPEX.  Both data sets are of excellent quality. The 70
$\mu$m-image, due to the lower angular resolution, shows only one
bright point-like source embedded in the globule. The image also shows
strong emission from the hot ionized rim facing $\sigma$ Ori and fainter
emission on the western side of the globule.

We have carried out multiframe PSF photometry using the SSC-developed
tool APEX on all the {\em Spitzer} IRAC images and on the  MIPS
images.  The 3.6 and 4.5~\micron\ IRAC images show signs of saturation
on bright stars.  We have used combination of automated routines and
eye-inspection to detect sources and extract photometry of these
sources from the IRAC and MIPS images.  For sources, which APEX failed
to detect at one or several wavelengths, we have used the user list
option in APEX to supply the coordinates for the source to extract it
and perform photometry on it. This enabled us to derive photometry for
every source, which we could visually identify on any image.

In the IRAC images we have detected 118, 122, 41 and 32 sources
respectively at 3.6, 4.5, 5.8 and 8.0~\micron.  Since the field of
view of the two IRAC cameras do not completely overlap, we end up with
a total of 125  mid-infrared sources. We detect five of them in the
24~\micron\ MIPS image, but only one in the  70~\micron\  image.  We
have crosscorrelated the sources detected in the IRAC and MIPS bands,
and the 2MASS point source catalogue. We have used the following
association radii : 1\arcsec\ for the IRAC images, 2\farcs5 for the
MIPS 24~\micron\ image and 2\arcsec\ for 2MASS data.
Table~\ref{tab_mirsrc} gives the coordinates of the 125 sources
identified in Ori\,I-2 together with their $R$-band and 2MASS
magnitudes, {\em Spitzer} IRAC and MIPS flux densities and a
preliminary classification based on selected color-color plots. In
Table~\ref{tab_mirsrc} we have given them the prefix Ori\,I-2, but
throughout the paper we simply refer to them as MIR-nn, where nn is
the number of the source.  Out of the 125 sources 51 were found to
have 2MASS counterparts. All the sources detected in the
MIPS~24~\micron\ image were also detected in the four IRAC bands. We
find that around 50\% of the sources in the list have been detected
only at 3.6 and 4.5~\micron\ and have no counterparts in 2MASS as
well, although many  of them are seen in the deep $R$-band image.

\subsection{Optical CCD imaging}
\label{NOT}

The  optical CCD images presented here were obtained on October 26
2007 using the 2.56~m Nordic Optical Telescope (NOT) located at El
Observatorio del Roque de Los Muchachos on the island of  La Palma in
the Canary Islands. We used the Andalucia Faint Object Spectrometer
and Camera (ALFOSC), which uses a thinned Loral 2048 $\times$ 2048 CCD
array with 15~$\mu$m pixels giving a field of view of 6.5 $\times$ 6.5
arcmin$^2$ with 0.187 arcseconds~pixel$^{-1}$. For these observations
we obtained two images with 300 seconds exposure time in Bessel R
(6500 \AA{}) separated by offsets of $\sim -$90\arcsec,+90\arcsec\ from the center of the globule in right
ascension and declination, and a short 30 second exposure centered on
the globule. We also observed the globule in narrow-band H$\alpha$. The
narrow band H$\alpha$ filter was a circular filter giving a more
restricted field of view and centered on 6564 \AA\  with a bandwidth of
33 \AA. Here we obtained two images, each with 300 second exposure and
separated by 60\arcsec\ in right ascension.  The observing conditions
were excellent with an average seeing of $\sim$ 0\farcs75.

The images were reduced with the STARLINK program suite CCDPACK. The
transformation to the World Coordinate System (WCS) was done using the
STARLINK program Gaia and we used more than twenty 2MASS stars as astrometric
reference stars. The 2MASS stars have very good astrometric accuracy
and the fit to these reference stars indicate that the astrometry of
the CCD images have an accuracy of $<$ 0\farcs1 for both the R and the
narrow-band H$\alpha$ image compared to the 2MASS reference frame.

The images of the photometric standard stars were unavailable, hence the
$R$-band image was calibrated using the 30 second exposure and seven
stars in the 14  - 15$^m$ range from the USNO-A2.0 catalogue. We
estimate the photometric calibration accuracy to be $\sim$ 0.1 -
0.2$^m$. This calibration was transferred to the H$\alpha$ image using a
somewhat larger set of overlapping stars with approximately neutral
colors judged from 2MASS J magnitudes. The H$\alpha$ image is estimated
to have a photometric accuracy no better than 0.3$^m$. The R band
image has a limiting magnitude of $\sim$ 23.5$^m$ and stars brighter
than 15$^m$ are partially saturated. We did aperture photometry in R and
H$\alpha$ of all stars, which coincided with an IRAC source to within
1\arcsec.  We present only the result of the R band photometry  in
Table~\ref{tab_mirsrc}. The  H$\alpha$ is virtually identical, except
for one star, see below. We detect all 2MASS sources in R band  as well
as 26 IRAC sources without 2MASS detection.

Comparison of the R and H$\alpha$ photometry revealed only one
H$\alpha$ emission line star. MIR-52 has an H$\alpha$ magnitude of
$\sim$ 18.5, while it is $\sim$ 19.9 in R (Table~\ref{tab_mirsrc}).
The excess seen in H$\alpha$, 1.4$^m$, is much larger than our
photometric uncertainty.  Therefore the star is definitely an
H$\alpha$ emission line star. MIR-52 has a strong IR excess in the
MIPS 24~\micron\ band, so that  we  identify it as a Class II object
based on IRAC and MIPS color color diagrams, see
Section~\ref{sec_ysoclass}.

\subsection{SCUBA observations}

The 850 $\mu$m and 450 $\mu$m continuum observations were obtained
with bolometer array SCUBA on JCMT\footnote{The JCMT is operated by
the Joint Astronomy Centre, on behalf of the UK Particle Physics and
Astronomy Research Council, the Netherlands Organization for
Scientific Research, and the Canadian National Research Council.},
Mauna Kea, Hawaii. SCUBA \citep{Holland99} has 37 bolometers in the
long and 91 in the short wavelength array separated by approximately
two beam widths in a hexagonal pattern. The field of view of both
arrays is $\sim$2.3\arcmin.  Both arrays can be used simultaneously by
means of a dichroic beamsplitter.

The SCUBA observations reported in this paper were all obtained in
jiggle-map mode \citep{Holland99}.  On October 17, 1997 we obtained
two separate five integration maps  with a chop throw of 100\arcsec\
in azimuth; which unfortunately was not sufficient to completely chop
outside of the globule.  On December 17, 1997  we therefore used a
120\arcsec\ chop with a fixed position angle of 80\degr\ measured from
North (equatorial reference frame).  This time we obtained three maps,
each with small position offsets (dither) with three, eight, and five
integrations. The total integration time from both nights was
therefore 3328 seconds.  The sky conditions were good on both nights.
On October 17 the atmospheric opacity measured with the Caltech
Submillimeter Observatory (CSO) taumeter, CSO $\tau_{\rm 225GHz}$  was
$\sim$ 0.07, and on December 17 $\tau_{\rm 225GHz}\sim$ was 0.04 -
0.05.  The dust emission from the globule is too extended to be
observed in jiggle map mode and the emission from the cold cloud core
is therefore likely to be underestimated, especially at 450 $\mu$m.
Any strong emission in the off source positions was carefully blanked
out in the data reduction stage.  It should not affect the morphology
or photometry of the strong, compact source, which dominates the
emission both at 850 and 450 $\mu$m.  Pointing corrections were
derived from observations of the blazar 0528+134 on October 17, and
the blazar 0529+075 on December 17. The secondary calibrators CRL\,618
and HL~Tau were used for flux density calibration. We estimate the
calibration accuracy to be $\sim$ 10\% at 850 \micron\ and $\sim$ 20\%
at 450 \micron. The Half Power Beam Width (HPBW) was estimated from
observations of Uranus obtained in the early part of each night and
was found to be $\sim$ 14\farcs5  -- 15\arcsec\ at 850 $\mu$m and
$\sim$ 8\arcsec\ -- 8\farcs5 at 450 $\mu$m.

The data were reduced in a standard way using SURF
\citep{Jenness99,Sandell01} and STARLINK imaging software, i.e., flat
fielded, extinction corrected, sky subtracted, despiked, and
calibrated the images in Jy {\rm beam$^{-1}$}. Each data set was
corrected for any drift in pointing between successive pointing
observations and the data were added together to determine the most
likely submillimeter position at 850 \micron. Once we had derived a
basic 850 \micron\ astrometric image, we made Gaussian fits of the
compact sub-millimeter source in each data set and derived small
additional RA and Dec corrections to each scan (shift and add) to
sharpen the final image to this position.  Since there is a small
mis-alignment between the 850 and 450 \micron\ arrays, we first
corrected the 450 \micron~images for any pointing drifts and then did
shift and add using the position derived at  850 \micron. We estimate
the astrometric accuracy  to be $\leq$ 2\arcsec.  The final coadd was
done by noise-weighting the data in order to minimize the noise in the
final images. The rms of the 450 \micron~image is $\sim$ 0.50  mJy
{\rm beam$^{-1}$} and $\sim$ 10 mJy {\rm beam$^{-1}$} for the 850
\micron\ image.  All the  maps were converted to FITS-files and
exported to MIRIAD \citep{Sault95} for further analysis.  In order to
correct for the error lobe contribution, especially at 450 \micron, we
have deconvolved all the maps using CLEAN with the same beam model
used by \citet{Hogerheijde00}. Since the HPBW varies slightly from
night to night, this model beam is not ideal, but it is the best we
can do. The actual beam size is probably somewhat more extended.

\begin{figure}[b]
\begin{center}
\includegraphics[angle=0,width=16.0cm,angle=0]{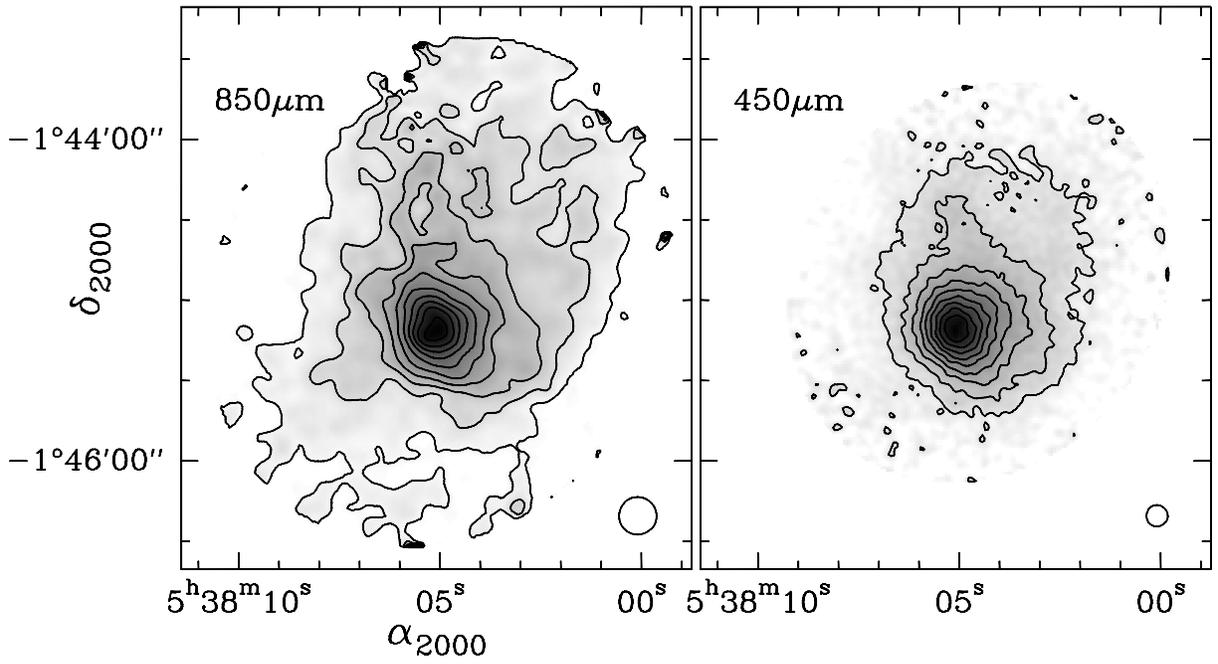}
\caption{Gray-scale 850 and 450~\micron\ continuum maps overlaid with
contours of Ori\,I-2. The peak intensities at 850 and 450~\micron\ are
0.21 and 1.39~Jy~beam$^{-1}$, respectively. The contours for both the 850 and
450~\micron\ maps are at 1\%, 10\% to 100\% of the respective peak
intensities in steps of 10\%.  
\label{fig_scuba}}
\end{center}
\end{figure}

Figure~\ref{fig_scuba} shows the  450 and 850~\micron\  SCUBA images
of Ori\,I-2. These images show a single bright source, located at the
head of the globule beyond the optically bright rim. This source,
SMM\,1,  is coincident with MIR-54, which is  the only source seen in
the MIPS 70~\micron\ image.  In order to derive the position and size
of SMM\,1 we have fitted a two component elliptical Gaussian using the
task IMFIT in MIRIAD, one for the sub-mm source, and the other for the
surrounding cloud.  The fit to the broader component is mainly to
provide a good subtraction of the extended emission, and is not to
estimate the flux density of the surrounding cloud.  The
sub-millimeter  position is  $\alpha$(2000.0) = 05$^h$  38$^m$
05\psec098, $\delta$(2000.0) = $-$ 01\degr\ 45\arcmin\ 11\farcs4. The
integrated flux densities are 0.54  $\pm$ 0.08 Jy and 3.36 $\pm$ 0.75
Jy for 850 and 450 \micron, respectively.  The fitted size,
17\farcs9$\times$ 15\farcs4, should be considered an upper limit,
since the true beam sizes were probably somewhat larger than the model
beams used for deconvolving the images. 


\section{The young stellar population in the Ori\,I-2 field}
\label{sect_ysopop}

Since Ori\,I-2 lies within the $\sigma$ Orionis cluster (age $\sim$ 3
Myr), we expect to detect some young stars belonging to the cluster
in the 5\arcmin\ $\times$ 5\arcmin\ field that we are analyzing. Stars
belonging to the   $\sigma$ Orionis cluster should be preferentially
located outside the globule and generally appear older than stars that
have recently formed in the globule.

\begin{figure*}
\begin{minipage}[t]{8.0cm}
   \includegraphics[width=8.0cm]{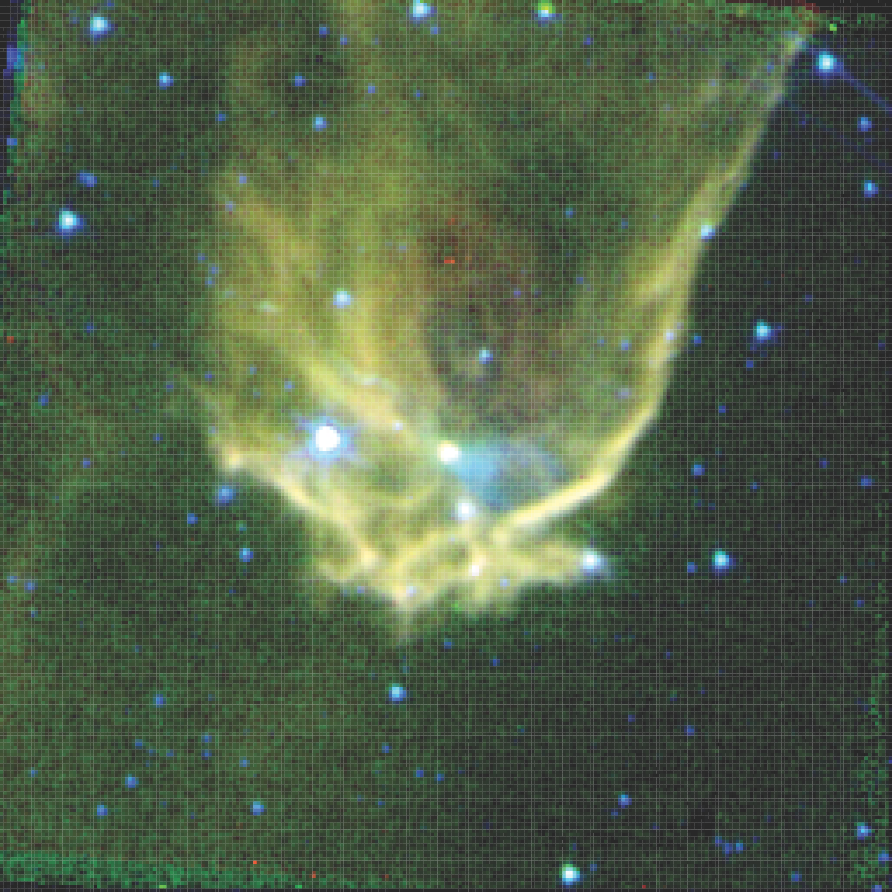}
  \end{minipage}
\begin{minipage}[b]{8.0cm}
   \includegraphics[width=8.0cm]{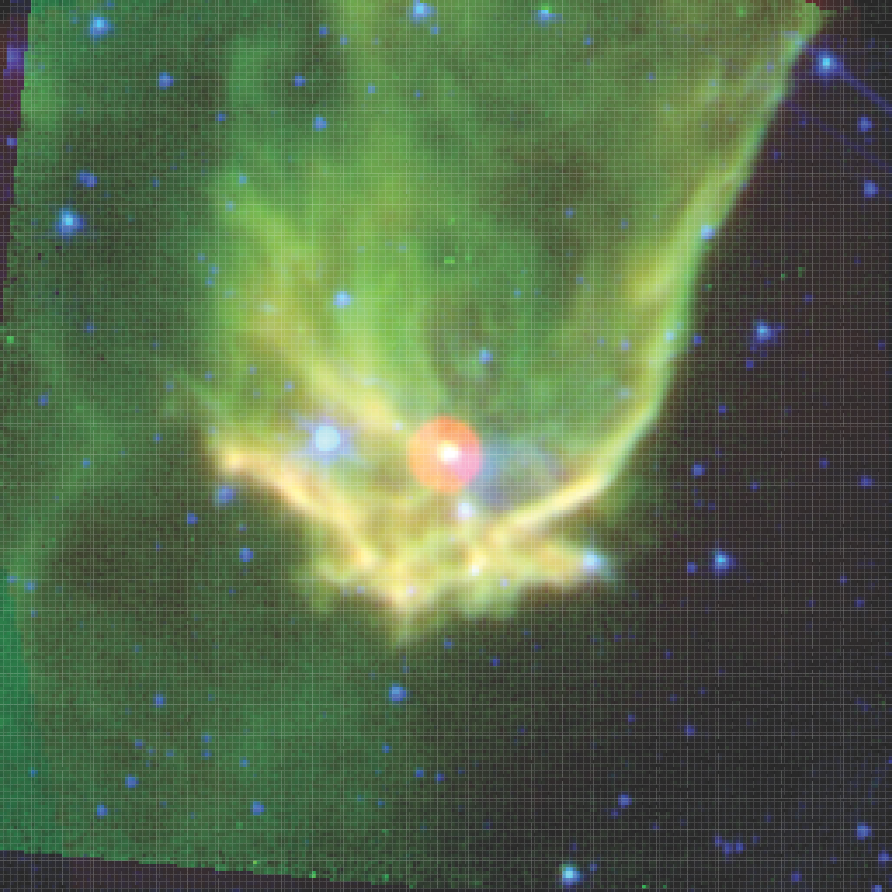}
  \end{minipage}
\caption{False-color {\em Spitzer} IRAC-MIPS images of Ori\,I-2. ({\em
Left}) Blue: 3.6~\micron; green: 5.8~\micron; red: 8.0~\micron. ({\em
Right}) Blue: 3.6~\micron; green: 8.0~\micron; red: 24.0~\micron.The
images are centered at $\alpha_{2000}$  = 05$^h$38$^m$05.3$^s$ and
$\delta_{2000}$ = -01\arcdeg45\arcmin10\arcsec, and extend over
5\arcmin$\times$5\arcmin\ ($\alpha\times\delta$).
\label{fig_composite}}
\end{figure*}

Figure~\ref{fig_composite} present three-color images of Ori\,I-2
using IRAC 3.6, 4.5 and 8 \micron\ in the {\em left} panel and the
IRAC 3.6 and 8 \micron\ bands combined with the MIPS 24  \micron\
channel in the {\em right} panel. Both images show that the front side
of the globule, which is dominated by strong polyaromatic hydrocarbon
emission (PAH-emission) is far from smooth. This may in part be due to
Rayleigh-Taylor instabilities at the ionization front, but it is also
due to outflows from the young stars which have recently formed in the
cloud. The outflow from MIR-54 (IRAS~05355-0146), the bright red star
in the IRAC/MIPS color image, is clearly seen as a cavity like
structure to the east of the star and a bluish nebulosity on the
western side of the star. This outflow is very prominent in the
H$\alpha$ image (Figure~\ref{fig_halpha_rband}), where it shows up as
a limb brightened cavity, extending $\sim$ 50\arcsec\ from the western
edge of the globule and $\sim$ 1.5\arcmin\ from the star exciting it.
Another outflow just south of the H$\alpha$ outflow and approximately
aligned with it, can be seen in the IRAC images terminating at a blue
star (unrelated to the outflow). This outflow is powered by another
young star, MIR-52, $\sim$ 20\arcsec\ to the south of MIR-54. To the
east both outflows overlap.  The  western side of the globule is very
sharp with an ionized rim. To the east the the boundary between the
globule on the surrounding \ion{H}{2} region is more diffuse. The most
opaque part of the globule is just north or northwest of the IRAS
source. The bright star located $\sim$ 43\arcsec\ away from MIR-54 in
the color image is SAO~132389 (HD~37389). This star, identified as
MIR-73 in Table~\ref{tab_mirsrc}, is detected even at 24~\micron,
although it is an unreddened foreground A0 star with an $m_{\rm V} =
8.3^m$.  The star has a radial velocity of $\approx 21$~km~s$^{-1}$,
which is very different from the radial velocity (12~\kms) of Ori\,I-2
\citep{cernicharo1992}. Most of the faint stars outside the globule
are field stars, although a few are young stars belonging to the
$\sigma$-Orionis cluster.

\subsection{YSO classification based on Near- and Mid-infrared colors
\label{sec_ysoclass}}

The IRAC and MIPS data are ideal for detecting young stellar objects;
especially in obscured regions like in Ori\,I-2, which has a visual
extinction exceeding 20$^m$. Outside the globule, where the extinction
is low, search for brown dwarfs, for which the Spectral Energy
Distribution (SED) peaks at J, can be most efficiently performed using
near-infrared data.  However, we go deeper than 2MASS in the 3.6 and
4.5~\micron\ IRAC bands. The same is true even for the NOT broadband R
filter. There is therefore no particular advantage in using the  2MASS
data alone. All 2MASS sources in our field have also been detected by
IRAC, at least in the two short wavelength bands. Furthermore, at the
distance of the $\sigma$ Orionis cluster  only the brightest M dwarfs
are barely detectable with 2MASS.

Since we can see several galaxies in the deep R band image, we first
analyzed the IRAC data to eliminate any galaxies, which could have MIR
colors similar to a YSO.  \citet{stern2005} demonstrated that (a) 
normal star-forming galaxies and narrow-line AGNs with increasing 5.8
and 8.0~\micron\ and (b) broad-line AGNs with red, nonstellar SEDs,
result in colors which are very similar to bona-fide YSOs. It is 
therefore necessary to inspect deep IRAC images for contamination due to
the extragalactic sources. \citet{gutermuth2008} have extensively
discussed the criteria for identifying such extragalactic objects in
the IRAC color-color diagrams. If we use the criteria given
in the appendix of their paper, we identify
MIR-16, 25, 88, 89, 93, and 114 as extragalactic. However,
inspection of these sources in the R-band CCD image, which was taken
in excellent seeing conditions, suggests that most of them are
stellar, with the exception of MIR-93. This source  is clearly extended, and
therefore it is an extragalactic source. For MIR-88, which is close to the detection
limit in the R-band, we get no help from the R band image, it could be
stellar, or it  could be extragalactic. MIR-89, which is not detected in the R band, 
was also identified as an
extragalactic object. However, not only IRAC, but also the MIPS colors
suggest that MIR-89 is a Class 0/I object. This classification is also
supported by its location in the eastern side of the bright rim.  We
therefore only positively identify MIR-93 as an extragalactic object.

\begin{figure*}[ht]
\begin{center}
\resizebox{\hsize}{!}{\includegraphics[angle=0,width=16.0cm,angle=0]{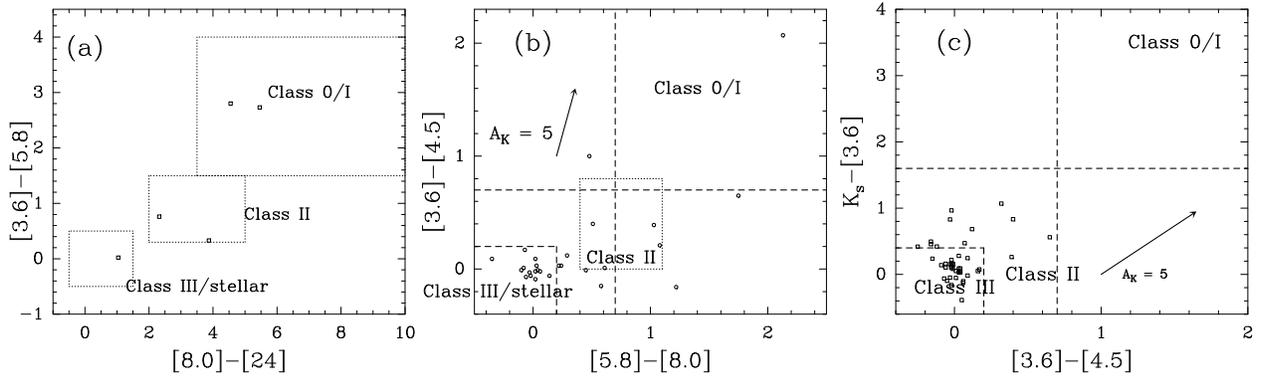}}
\caption{Color-color diagrams for all MIR and PMS-NIR sources.
Approximate classification criteria adopted from
\citet{megeath2004} and \citet{muzerolle2004} are shown respectively
for the left and the middle panel. (a) MIR color-color diagram based
on {\em Spitzer} 3-band IRAC and MIPS photometry. (b) 4 band IRAC
color-color diagram and (c) NIR-MIR color-color diagram based on 2MASS
$K_{\rm s}$ and two {\em Spitzer} IRAC bands.  Dashed lines in all
panels are taken from \citet{hartmann2005}. Reddening vector
corresponding to the extinction laws given by fitted function from
\citet{indebetouw2005} is shown in {\em (b)}. The dashed lines $K_{\rm
s}$--[3.6] = 1.6, [3.6]--[4.5] = 0.7, [4.5]--[5.8] = 0.7 and
[5.8]--[8.0] = 0.7 discriminate Class II sources from Class I/0
sources, and $K_{\rm s}$--[3.6] = 0.4, [3.6]--[4.5] = 0.2,
[4.5]--[5.8] = 0.2 and [5.8]--[8.0] = 0.2 discriminate Class III from
Class II. 
\label{fig_iracmipscol}}
\end{center}
\end{figure*}

Figure~\ref{fig_iracmipscol} presents  color-color diagrams of 2MASS,
IRAC and MIPS 24~\micron\ sources detected in the Ori\,I-2 field. We
have used several criteria (shown as dashed lines and boxes in
Figure~\ref{fig_iracmipscol}) to identify potential PMS stars using
these color-color diagrams.

The most stringent classification scheme uses the IRAC ([3.6]--[5.8])
colors and the [8]--[24] IRAC and MIPS color. At 24 \micron\ the
reddening due to extinction is small and the photospheric colors are
very close to zero for all spectral types \citep{muzerolle2004}.
Therefore the [8]--[24] color is very sensitive to infrared excess,
but of course not all young stars bright are enough to be detected at
24 \micron. Using this color-color diagram we find  four sources with
infrared excess (Figure~\ref{fig_iracmipscol} ({\em left})).  We
identify MIR-54  and MIR-89 as  Class 0/I sources, while MIR-41 and
MIR-52 are in the Class II regime.

Figure~\ref{fig_iracmipscol}~({\em middle}) presents the [3.6]--[4.5]
vs
[5.8]-[8.0] color-color plot for the sources detected in all the IRAC
bands.  Sources with the colors of stellar photospheres are centered
at ([3.6]--[4.5],[5.8]--[8.0])=(0,0) and include foreground and
background stars  as well as diskless (Class III) pre-main sequence
stars. The box outlined in Figure~\ref{fig_iracmipscol} ({\em
middle}), defines the location of Class II objects
\citep{megeath2004,allen2004}, i.e. sources whose colors can be
explained by young, low-mass stars surrounded by disks.
\citet{hartmann2005} have shown from their observations of young stars
in the Taurus-Auriga complex that  Class 0/I protostars require
[3.6]--[4.5]$>0.7$ and [5.8]--[8.0]$>0.7$. While MIR-89
clearly falls in the Class 0/I region, MIR-54 does not. It is well
above the horizontal line corresponding to [3.6]--[4.5]=0.7, but the
color
[5.8]--[8.0] is slightly less than 0.7. Since the demarcating lines
primarily serve as a guidance and MIR-54 is securely identified as a
Class 0/I object based on the IRAC/MIPS color-color plot, we identify
it as a Class 0/I object here as well. In this color-color diagram
five stars end up in the Class II color regime: MIR-23 (marginally),
MIR-25, MIR-41, MIR-51, and MIR-88.  MIR-51, which shows a large
excess in the 
[5.8]--[8.0] color is  a Class II/I source, i.e. a flat spectrum
source.
We did not pick up MIR-52, because the [3.6]--[4.5] color is only
0.12, yet it is the only strong H$\alpha$ emission line star in our
sample (see Section~\ref{NOT}) and definitely a PMS star. It has,
however, a strong infrared excess longward of 8~\micron, which is why
it was picked up in the [3.6]--[5.8] vs.  [8]--[24] color-color
diagram.

\citet{hartmann2005} also investigated color-color diagrams using the
shortest IRAC bands, which have the highest sensitivity and which are
least affected by PAH emission with near-IR colors taken from the 2MASS
catalogue. We have similarly used the photometry for the sources
presented in Table~\ref{tab_mirsrc} which have 2MASS counterparts to
plot the $K_{\rm s}$--[3.6] versus [3.6]--[4.5] color-color diagram
(Figure~\ref{fig_iracmipscol} {\em right}). We identify 12 potential PMS
stars, i.e. stars which are located outside the box demarcating stellar
sources. These sources are MIR-10, 29, 41, 47, 49, 51, 52, 61, 63, 78,
93,  and 117. If we account for the large errors in faint 2MASS sources
we find at least one more source, MIR-17, which appears to have a clear
excess in the 5.8~\micron\ filter. For both MIR-10 and 78 the 2MASS
$K_{\rm s}$ fluxes have large uncertainties. It is therefore desirable
to have additional evidence of IR excess in order to justify their
classification as PMS objects. While MIR-10 shows a clear excess at
5.8~\micron, MIR-78 is not detected beyond 4.5~\micron\ and it's SED is
completely stellar with no sign of an IR excess. Thus we do not consider
MIR-78 to be a PMS object. Since MIR-93 was found to be an extragalactic
source, we are left with 10 potential PMS stars. Of these we have
already identified MIR-41, 49, 51 and 52 as PMS stars based on MIR
color-color plots. The two Class 0/I sources, MIR-54 and 89,  and
MIR-88, a potential Class II source, were not detected in $K_{\rm s}$,
and hence do not appear in this color-color diagram. Neither do the
sources MIR-23, and 25, both of which were classified as PMS objects in
the IRAC color-color diagram, show $K_{\rm s}$--[3.6]  or [3.6]--[4.5]
colors expected for Class II objects. The $K_{\rm s}$--[3.6] color is
0.3 for MIR-23, which would put it in the class II regime, while it is
$\sim$ 0 for MIR-25. The latter, however, does have a significant color
excess at 8~\micron.

Not all of the stars identified in the 2MASS/IRAC color color diagram
are likely to be PMS stars, they could just be heavily reddened
background stars. MIR-29, 47, and 61 are all located in the western or
southern rim of the globule, and have very blue IRAC colors. They are
all almost certainly heavily reddened background stars. The same is
true for MIR-63  which is seen toward the opaque part of the globule.
We identify MIR-63 with the star that \citet{cernicharo1992} labeled
B, and proposed might be a stellar counterpart to the IRAS source.
However, the [3.6] -- [4.5] color for MIR-63 is extremely blue,
$-$2.4$^m$, which shows that the red near-IR colors are simply due to
extinction. The same is true for MIR-117, which is located east of the
globule in a relatively low extinction area. For this star the $K_{\rm
s}$--[3.6] color excess is rather marginal and there is no excess in
the IRAC bands.  It appears to be a reddened background star. MIR-10
lies outside the globule in a region of low extinction and is very red
in the [4.5] -- [5.8] color band, and could therefore be a PMS star,
although it was not detected at 8~\micron.

In all we therefore identify 11 potential PMS objects based on 2MASS,
IRAC and MIPS colors. We classify MIR-54 and 89 as Class 0/I stars ,
while MIR-10, 17, 23, 25, 41, 49, 51, 52, and 88 are Class II objects.
We also note that MIR-104, which lies just outside the southern rim of
the globule, was only detected at 5.8 and 8~\micron, suggesting that it
has an infrared excess. It is therefore a potential PMS object.  Among
these PMS objects MIR-10, 17, and 41 are located well south of the
globule, while MIR-25 is west of the cometary globule. If all of these
four sources are PMS stars, they most likely belong to the young
$\sigma$-Orionis cluster and are unrelated to Ori\,I-2. All the rest
could potentially have formed in the globule.  

In Section~\ref{sed_modeling} we investigate the SEDs of these
candidate PMS objects in more detail to see whether they are young stars,
and if so, what their physical properties are.

\subsection{The large Herbig-Haro outflow, HH\,289,  and a new HH object, HH\,992}
\label{sec_hh992}

\citet{mader1999} identified a parsec scale Herbig-Haro outflow,
HH\,289, to the east of the globule by analyzing images from the
AAT/UKST H$\alpha$ survey, broadband R (IVN) plates and narrowband CCD
images in H$\alpha$ + [\ion{S}{2}].  Their HH objects 2 -- 5, i.e. knots
C -- F, are all outside the area we imaged in R and H$\alpha$.  We do
detect their knot B in our H$\alpha$ image, but not knot A. Knot B appears
very elongated and is located 
in the cavity wall from the red-shifted outflow from MIR-54.
\citeauthor{mader1999}'s feature 1, which coincides
with the outflow cavity emerging to the SW out of the globule is also readily visible in the
H$\alpha$ image (Figure~\ref{fig_halpha_rband}). Although we see no trace of knot A in
either R or H$\alpha$, it seems to coincide with the bluish nebulosity
west of MIR-54, which is very prominent in the IRAC three color image
(Figure~\ref{fig_composite}). In the 3.6 and 4.5~\micron\ images the
MIR-54 nebulosity appears to be the inner part of an outflow cavity
which is not visible in the mid-IR outside the bright western rim of the
globule. However, this outflow lobe is seen more prominently as a
limb-brightened H$\alpha$ cavity extending out another 50\arcsec\ from
the western edge of the globule.  The total length of the outflow is
therefore $\sim$ 1.5\arcmin, 0.17 pc. Since this outflow extends into
``empty'' space, i.e. the low density \ion{H}{2} region IC\,434, it is
possible that the outflow extends even further, but becomes so diffuse
that it is no longer visible. It is probably the counterpart to the
large Herbig-Haro flow, but since the density of the globule is much
lower on the eastern side of the globule, it has been able to expand
more freely. However, in the IRAC images one can also see an outflow
cavity from MIR-52 protruding out of the globule. This outflow is
approximately parallel with the MIR-54 outflow. It is brighter on the
southern side of the cavity and has an apparent length of $\sim$
50\arcsec\ from MIR-52. To the east the two outflows appear as a large
evacuated hole (cylinder) and cannot be separated. Either star could
therefore drive the large HH outflow.

However, we also detect a Herbig-Haro like nebulosity with three knots
or condensations in the R-band image $\sim$ 140\arcsec\ east of MIR-54.
A blow-up of this nebulosity is shown in Figure~\ref{fig_halpha_rband},
although it is not seen at all in  the H$\alpha$ image. Normally HH
objects are bright in H$\alpha$. If this is an HH object, the emission
must completely be dominated by [SII] emission, i.e. it has to be an
extreme low excitation HH object. It is not detected by 2MASS, nor is it
seen in the IRAC images, which go deeper than 2MASS. It is, however,
seen faintly in \citeauthor{mader1999}'s H$\alpha$ + [SII] image, which
proves that it is real and not just an instrumental artifact. In fact it
appears quite strong in \citeauthor{mader1999}'s 2.12 \micron\ molecular
hydrogen image, while the nebulosity is completely absent on their  IVN
plate. This confirms that there is no continuum at the position of the
nebulosity. The only conclusion we can draw, is that
this is a rather low excitation HH object, which is  completely
dominated by line emission from  [SII] and which is also seen in
vibrationally excited H$_2$ at 2.12 \micron. The coordinates of the
three HH knots, which are here labeled HH\,992, are given in Table~\ref{tab_hhobjects}. This HH object is
not a part of the large Herbig-Haro flow, HH\,289, seen by  \citet{mader1999}. The
most likely exciting source of the HH\,992 complex is MIR-89, which is
$\sim$75\arcsec\ west of the HH objects. MIR-89 was identified as a
Class 0/I object in the color-color diagrams, but SED modeling
(Section~\ref{sed_modeling}) suggest that it is a Class II object. In
the 3.6 and 4.5~\micron\ images one can see a limb-brightened fan-shaped
nebulosity emerging to the west from the star. This nebulosity may trace
an outflow into the globule, which is roughly aligned (p.a. $\sim$
-95\degr{}) with the HH objects on the opposite side of the star. 

\section{Discussion of young objects in Ori\,I-2}

\subsection{Detailed SED modeling of the PMS objects}
\label{sed_modeling}

In order to interpret the observed SEDs for the sources identified to be
PMS and better characterize these sources we have explored the archive
of two-dimensional (2D) axisymmetric radiative transfer models of
protostars calculated for a large range of protostellar masses,
accretion rates, disk masses and disk orientations created by
\citet{robitaille2007}.  This archive also provides a linear regression
tool which can select all model SEDs that fit the observed SED better
than a specified $\chi^2$.  Each SED is characterized by a set of model
parameters, such as stellar mass, temperature, and age, envelope
accretion rate, disk mass, and envelope inner radius.  We have used this
online tool to generate models, which fit the observed SEDs for the 11
candidate PMS objects.  SEDs of six of the objects could be adequately
fit with accretion disk models, the other five were better fit with
stellar atmospheric models, reddened by foreground extinction. We
restricted the SED fitting tool to explore only distances between 350
and 450~pc. Below we present and briefly discuss the best fit models.

\begin{figure*}[htbp]
  \centering
  \begin{minipage}[b]{5.0cm}
    \includegraphics[width=5.0cm]{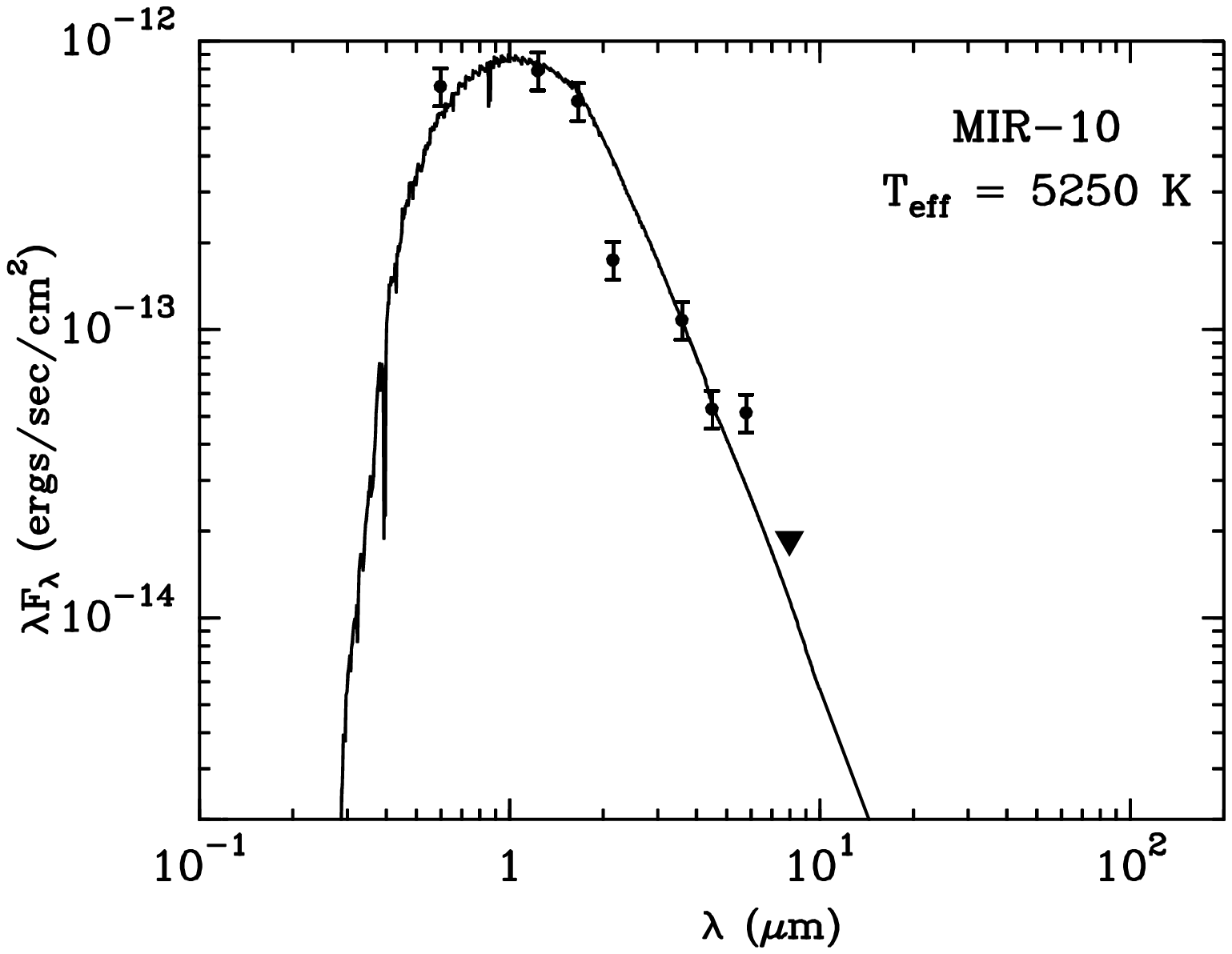}
  \end{minipage}
  \begin{minipage}[b]{5.0cm}
    \includegraphics[width=5.0cm]{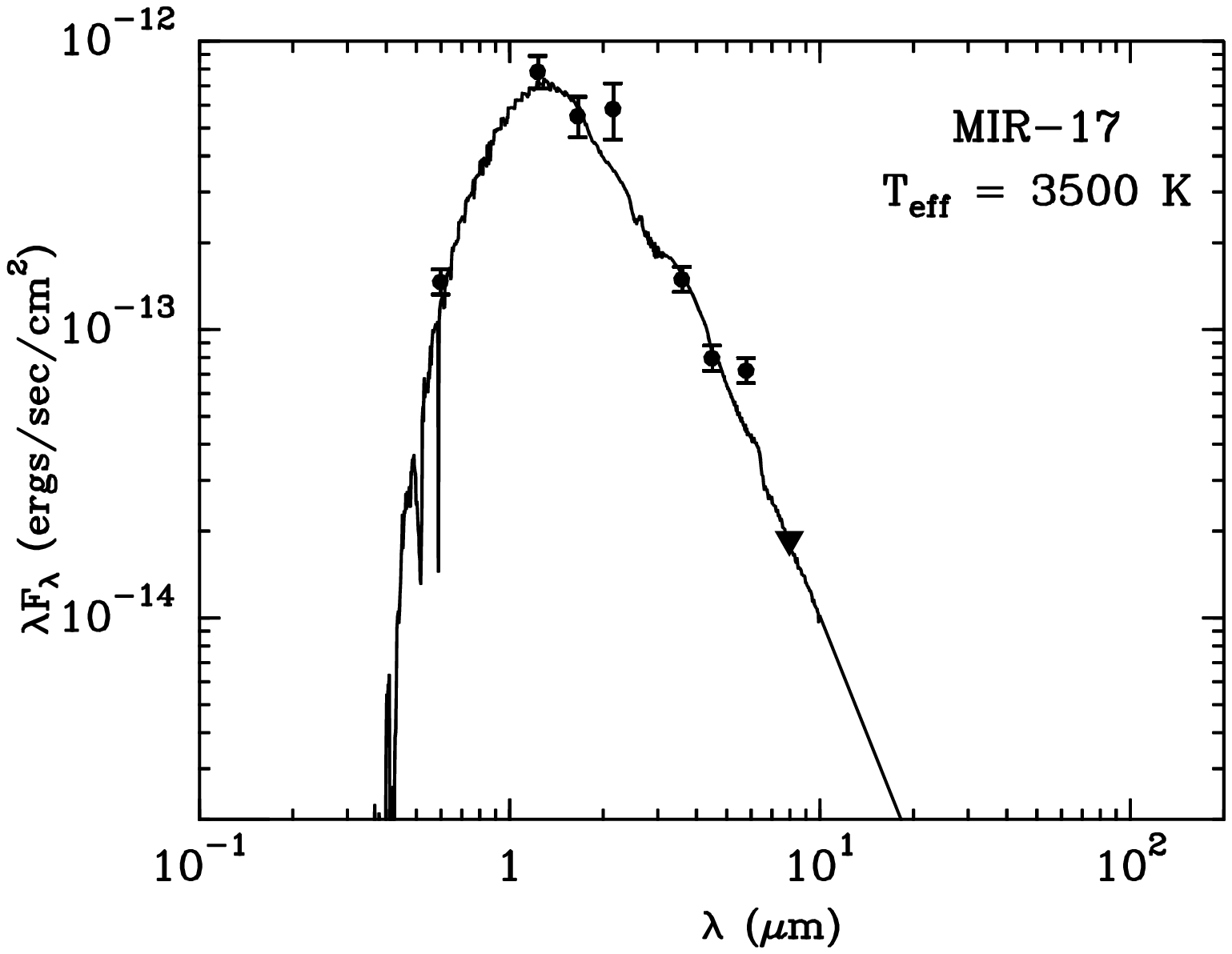}
  \end{minipage}
  \begin{minipage}[b]{5.0cm}
    \includegraphics[width=5.0cm]{mir23sed_multi.eps}
  \end{minipage}
  \vspace{0.5cm}
  \begin{minipage}[b]{5.0cm}
    \includegraphics[width=5.0cm]{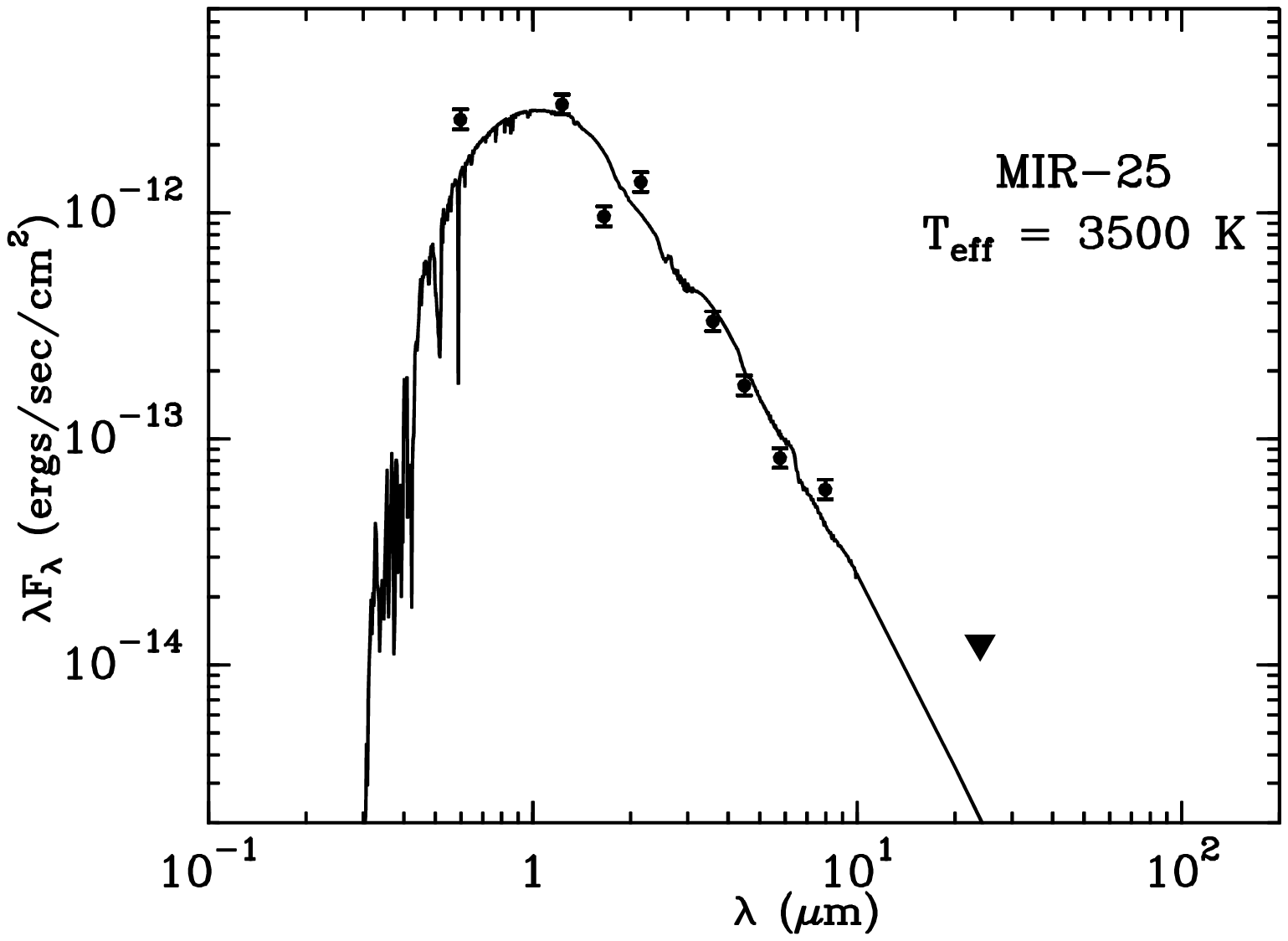}
  \end{minipage}
  \begin{minipage}[b]{5.0cm}
    \includegraphics[width=5.0cm]{mir41sed_multi.eps}
  \end{minipage}
 \hspace{0.2cm}
  \begin{minipage}[b]{5.0cm}
    \includegraphics[width=5.0cm]{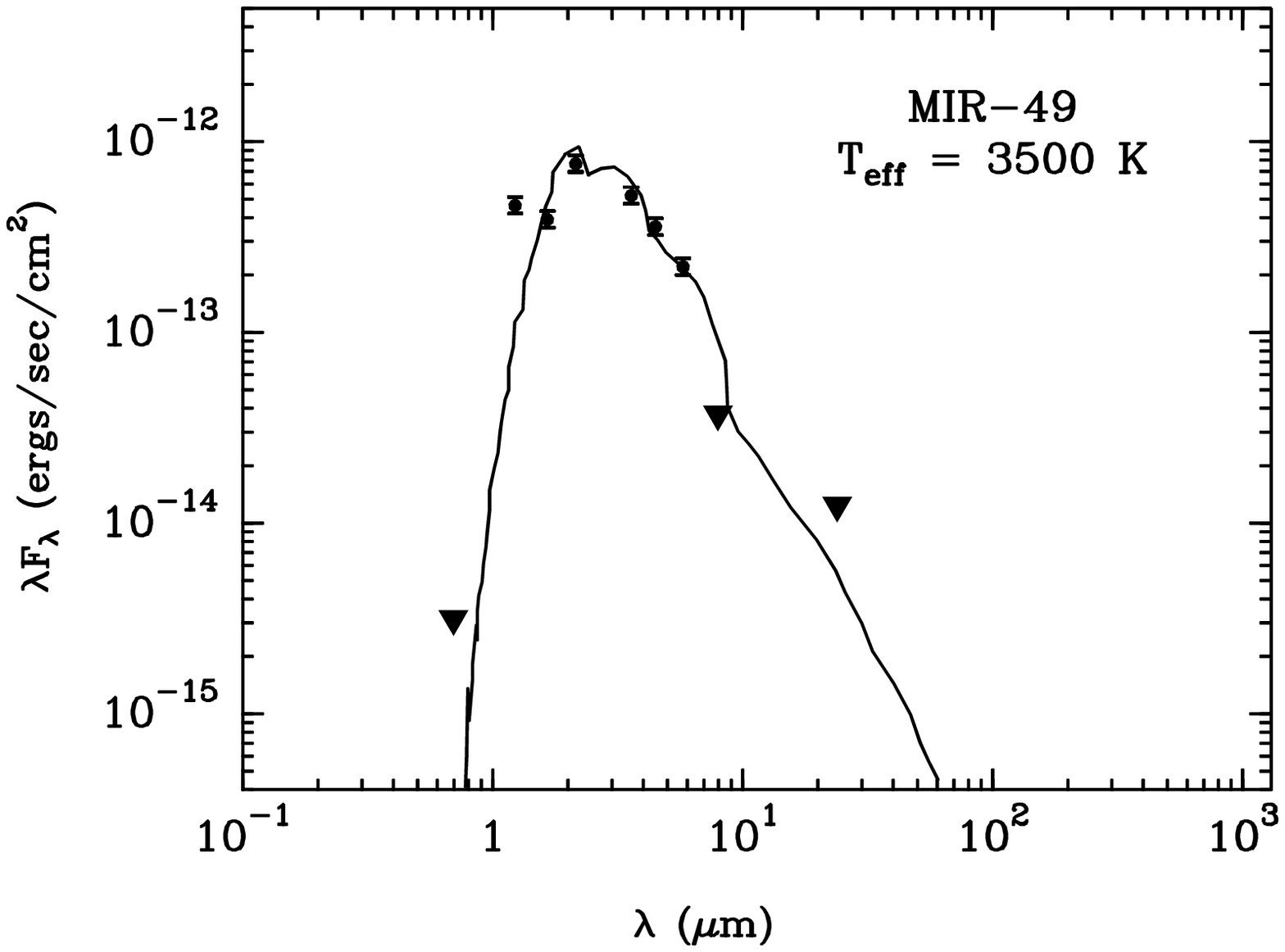}
  \end{minipage}
  \vspace{0.5cm}
  \begin{minipage}[b]{5.0cm}
    \includegraphics[width=5.0cm]{mir51sed.eps}
  \end{minipage}
  \begin{minipage}[b]{5.0cm}
    \includegraphics[width=5.0cm]{mir52sed.eps}
  \end{minipage}
  \begin{minipage}[b]{5.0cm}
    \includegraphics[width=5.0cm]{mir54sed.eps}
  \end{minipage}
  \vspace{0.5cm}
  \begin{minipage}[b]{5.0cm}
    \includegraphics[width=5.0cm]{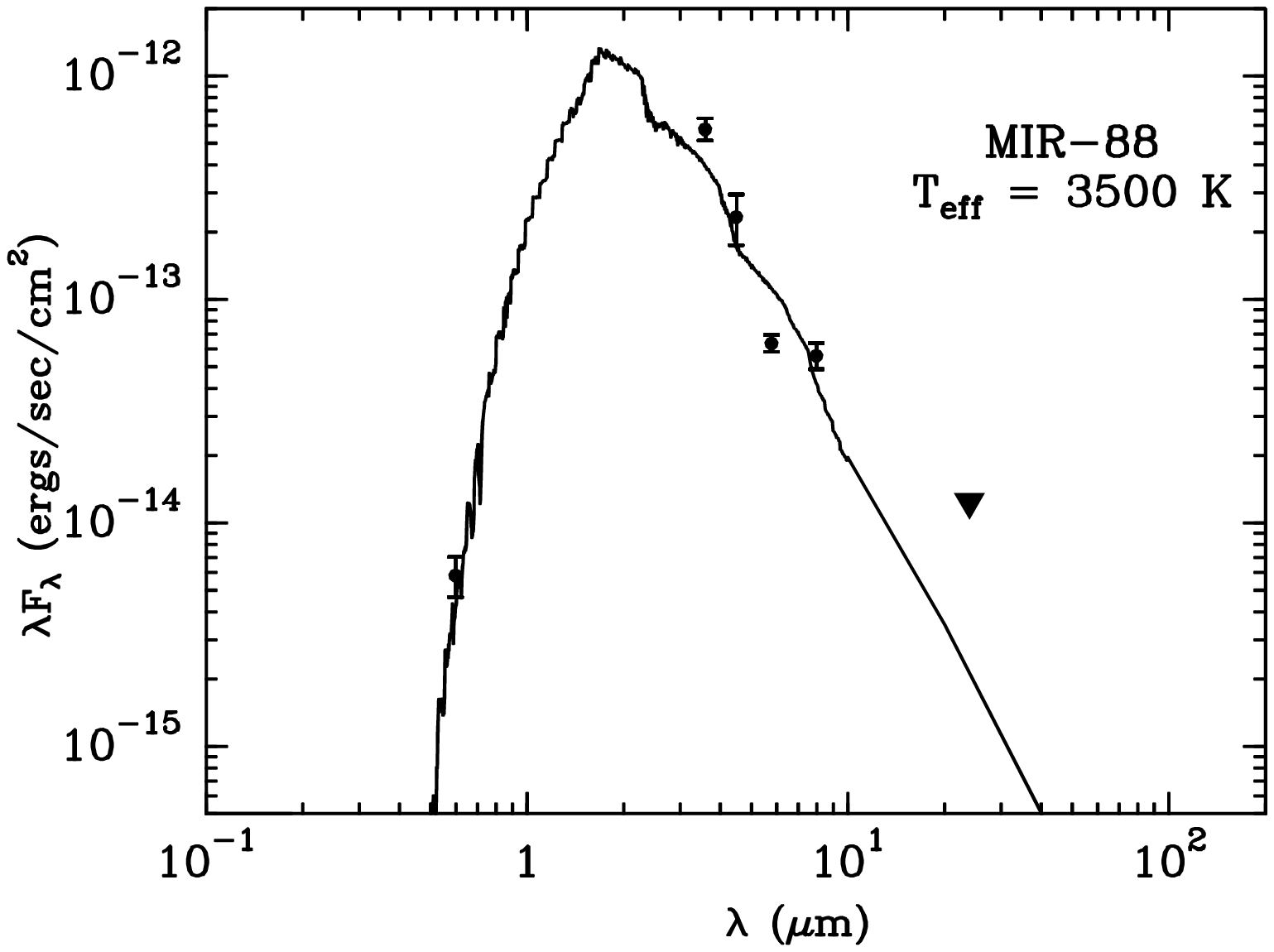}
  \end{minipage}
  \begin{minipage}[b]{5.0cm}
    \includegraphics[width=5.0cm]{mir89sednew.eps}
  \end{minipage}
\caption{SED fits for the PMS objects using axisymmetric radiation
transfer models and stellar photospheres (for MIR-10, 17, 25, 49 and
88). The filled circles indicate the measured fluxes and
uncertainties. The filled triangles correspond to the upper limits of
flux at these wavelengths.  The continuous black line represents the
best-fitting SED, the grey line (where present) represent other
acceptable models and the dashed line shows the stellar photosphere
corresponding to the central source of the best fitting model, as it
would look in the absence of circumstellar dust (but including
interstellar extinction).  For the SEDs fitted with stellar
photospheres, the stellar temperatures \tstar\ are also indicated
within the panel. 
\label{fig_sedfits}}
\end{figure*}

Figure~\ref{fig_sedfits} shows results of detailed modeling of the
observed SED in the mid-infrared (and FIR and sub-mm for MIR-54) for
all the candidate PMS objects.   A major criticism against the
use of these models has been the non-uniqueness of the solutions
obtained from the model library. However we find that for four out
of the six PMS objects, {\em viz., MIR-51, 52, 54 and 89} the best
fit models are either distinctively better in reproducing all the
observed flux densities than the next few models or that the distances for
other possible models do not match the distance to Ori\,I-2. For MIR-23 
we find 14 models with $\chi^2$--$\chi^2_{\rm
best} < 10$ and for MIR-41 we find four  models with
$\chi^2$--$\chi^2_{\rm best} < 2$, which appear plausible. For MIR-23 we
find two independent models for which the inclination is
varied to generate the 14 models.  For each of these two sources we
have obtained the average values and variances of the fitted
parameters by taking a weighted average of the apparently acceptable
models in a manner similar to \citet{simon2007}.
Table~\ref{tab_sedfits} presents the parameters corresponding to the
best fit models for the six sources which we fitted with accretion disk models.

\citet{robitaille2006} presented a classification scheme which is
essentially analogous to the Class scheme, but refers to the actual
evolutionary stage of the object based on its physical properties like
disk mass and envelope accretion rate rather than the slope of its
near/mid-IR SED.  According to the \citeauthor{robitaille2006}
classification scheme, MIR-54, is a stage 0/I object, while all the
others are stage II objects. At a first glance it is surprising to
find that MIR-89, which we classified as a young Class 0/I object both
in the IRAC and IRAC-MIPS color-color diagrams, is now identified as
an evolved stage II object. The SED model shows that it only has  a
minuscular envelope and there is no sign of envelope accretion.  All
reasonable SED models suggest that it is a heavily extincted (\av\
$\geq$  50$^m$), very low-mass star. This is the sole reason why it
appears as a Class 0/I object in the color-color diagrams, because the
observed magnitudes are not corrected for extinction in these
diagrams.  The star still has a rather massive disk compared to the
stellar mass, $M_{\rm disk}/$\mstar $\sim$ 0.02, and a moderately high
disk accretion rate of 3.2~10$^{-8}$~\msun~yr$^{-1}$.  It is located
in the bright rim on the eastern edge  of the cometary globule
Ori\,I-2, and appears to drive an outflow seen as a fan-shaped
nebulosity west of the star, and as the Herbig-Haro objects, HH\,992,
seen $\sim$65\arcsec, 0.12~pc to the east of the star.

MIR-51, which we identified as a Class I/II object based on color-color
diagrams, appears more similar to MIR-52 (Class II object) and if
anything slightly older, not younger. Both appear to be young stage II
objects (${\dot{M}}_{\rm env}$/\mstar $>$ 10$^{-7}$ yr$^{-1}$ and
$M_{\rm disk}$/\mstar\ $>$ 10$^{-4}$) and are embedded inside the globule.
This is consistent with MIR-52 having strong H$\alpha$ emission and
powering an outflow.

The SED models place MIR-23 and 41 as late stage II objects. They have
no measurable envelope and very low disk accretion rates
(Table~\ref{tab_sedfits}).  The models also suggest no or  low
foreground extinction. These results of modeling together with the
location of these two objects in the region suggest that these objects
are therefore more evolved PMS stars belonging to the $\sigma$-Orionis
cluster, although we cannot exclude that they could have formed in the
globule at an earlier epoch.

We could not obtain any believable dust/envelope models for MIR-10,
17, 25, and 88, which are all outside the globule in regions of low or
moderate extinction, or for MIR-49, which is located in the most
opaque part of the globule.  We therefore fitted these objects with
pure stellar atmospheric models reddened by foreground extinction.
These fits are not very accurate, since we have no optical data, other
than the R magnitudes. Furthermore, MIR-10, 17 and 49 were not
detected at 8 \micron, while MIR-88 was not detected by 2MASS. We have
however used upper limits from these bands to better constrain the
fits.  {\bf MIR-10:} At wavelengths shortward of 4.5\micron, the
observed SED is fitted reasonably well by a stellar photosphere \tstar
= 5250~K, log$[$Z/H$]$=-0.5, log$[$g$]$ = 0.0 and \av = 1.45$^m$.
However, the star shows an excess at 5.8~\micron, which suggests that
it could be a transitional disk object. {\bf MIR-17:} Similar to
MIR-10 the 2MASS and short wavelength IRAC data for this star can be
explained by reddening.  Even though it also shows an IR excess at
5.8~\micron, when compared with the SED due to stellar photosphere
\tstar = 3500~K, log$[$Z/H$]$=-0.5, log$[$g$]$ = 5.0 and \av $\sim$
2$^m$, the 8~\micron\ upper limit suggests that there is no excess at
8~\micron. We therefore dismiss MIR-17 as a PMS star. {\bf MIR-25:} A
stellar model with $T_\ast$ = 3500~K, log $[$Z/H$]$=-2.5, log$[$g$]$ =
5.0,   and \av = 0$^m$ gives a reasonable fit to the observed data.
There might be a small excess at 8~\micron, which could continue to
longer wavelengths.  Based on the available data, we therefore retain
MIR-25 as a candidate PMS star, although the evidence for dust excess
in this star is rather marginal. {\bf MIR-49:} The SED of this source
is well reproduced by a stellar photosphere with $T_\ast$ = 3500~K and
\av = 16$^m$. MIR-49 therefore appears to be a heavily reddened
background star. {\bf MIR-88:} The stellar model with  \tstar =
3500~K, log$[$Z/H$]$=-0.5 , log$[$g$]$ = 0.0 \av=6.1$^m$ reproduces
the observed SED significantly better than the accretion disk models
and there is no evidence for excess emission in the IRAC bands. It is
therefore a reddened background star.

Most of the stars, which could not be fit with accretion
disk models, appear to be reddened background stars, except perhaps for
MIR-10 and 25, both of which could have some long wavelength excess.
It is possible that they could be transitional disk objects belonging to the
$\sim$ 3 Myr $\sigma$ Orionis cluster. Transitional disk objects are
stars,  for which the inner disk has been cleared out, see e.g.
\citep{forrest2004}, and which therefore do not show any excess until
6 or 8~\micron. 
 
For MIR-54 the models completely fail to reproduce the observed
sub-millimeter flux densities, irrespective of the values of any of the
input parameters.  We discuss MIR-54 in more detail in
Section~\ref{sect_MIR-54}.

Based on our SED modeling  we conclusively identify  six PMS stars.
All the PMS stars in the Ori\,I-2 region appear to be low luminosity
stars of spectral types K and M. The PMS stars outside
the  globule have very low disk
masses and disk accretion rates, 10$^{-10}$ to a
few times 10$^{-11}$~\msun~yr$^{-1}$ (Table~\ref{tab_sedfits}).  This
is not surprising since these stars are most
likely members of the $\sigma$-Orionis cluster, which has an age of
$\sim$ 3~Myr \citep{caballero2008}.  

\subsection{MIR-54, a low-mass Class 0/I object
\label{sect_MIR-54}}

MIR-54 is the mid-infrared counterpart to IRAS~05355-0146, which was
detected in all four IRAS bands. MIR-54 is within 13\arcsec\ of the
nominal IRAS position. It is the only source in Ori\,I-2 detected at
24 and 70 \micron\ as well as in the sub-millimeter. The MIPS
70~\micron\ flux density for MIR-54 is $3.28\pm0.01$~Jy. The IRAS flux
densities are much higher than what we observe with MIPS.  This is to
be expected, since the large IRAS beam will include emission from the
hot bright rim as well.  The luminosity for the IRAS source, i.e.
MIR-54, has therefore been severely over estimated in the past.

\begin{figure}[ht]
\begin{center}
\resizebox{\hsize}{!}{\includegraphics{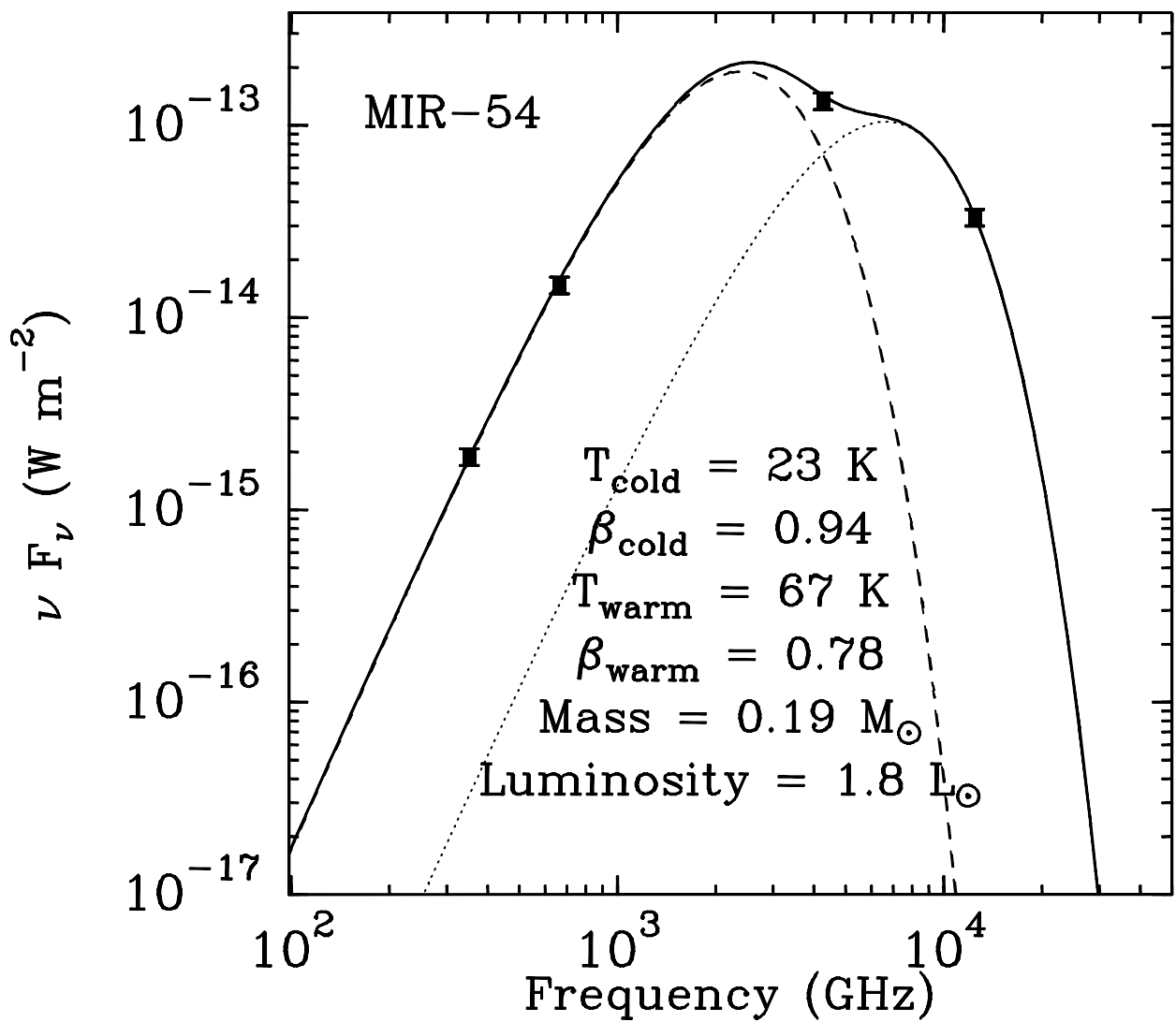}}
\caption{Two-temperature greybody fit to the SED of MIR-54.  The
fitted dust temperature, \tdust, and the dust emissivity index
($\beta$)  for the warm and cold components and the total mass and
luminosity for the best fit model are written in the figure.
\label{fig_greyfit}}
\end{center}
\end{figure}

We can obtain a more precise estimate of the bolometric luminosity by
doing a greybody fit to the MIPS and SCUBA data.  This is illustrated
in Figure~\ref{fig_greyfit} where we show a two-component
greybody fit to the observed flux densities between 24 and
850~\micron.  Based on the observed size of the source in the
mid-infrared and sub-mm, we have assumed $\theta_s$ to be 10\arcsec\
and 13\arcsec\ for the warm and the cold components, respectively. The
dust temperatures (\tdust), dust emissivity indices ($\beta$) and
masses calculated for the best fit model are presented in
Figure~\ref{fig_greyfit}. The total dust and mass gas, $M_{\rm tot}$ was
derived from the fitted \tdust\ and $\beta$ using

\begin{equation}          
M_{tot} = 1.88 \times 10^{-2} \biggl({1200\over\nu}\biggr)^{3+\beta}S_\nu
(e^{0.048\nu/T_d}-1)d^2,
\end{equation}

where $d$ is the distance in $[$kpc$]$, S$_\nu$ is the total flux (in
Jy) at frequency $\nu$ and  M$_{\rm tot}$ is given in $[$\msun$]$.  

Based on the greybody fitting we obtain dust temperatures of 23~K and
67~K and $\beta$ of 0.9 and 0.8 for the cold and the warm components.
Between 24~\micron\ and sub-mm wavelengths, the observed SED of the
source is reasonably well fit using this simple two-temperature
greybody model and it is consistent with the observed 1.3~mm flux
density.  However we could not derive a single radiative transfer
based model, which reproduces the entire SED starting from
3.6~\micron\ to the sub-mm wavebands (Figure~\ref{fig_sedfits}). The
model is particularly poor at the longer wavelengths. However,
both the greybody model and the radiation transfer based models give a
luminosity of 1.3--1.8~\lsun\ and mass of 0.19--0.24~\msun.  MIR-54 is
therefore a deeply embedded (not detected in the near-IR) low-mass, very
young object. It excites an
H$_2$O maser \citep{wouterloot1986,codella1995}.  It is most likely the exciting source for the large
parsec scale outflow, HH\,289. Inside the globule this is seen as a low
velocity molecular outflow \citep{cernicharo1992}. The outflow is also visible
in the IRAC images (Section~\ref{sec_hh992}). The SED modeling
predicts ${\dot{M}}_{\rm env}$/\mstar $\sim$ 2~10$^{-5}$ yr$^{-1}$,
which satisfies the criterion (${\dot{M}}_{\rm env}$/\mstar  $>$
10$^{-6}$ yr$^{-1}$) used by \citet{robitaille2006} to identify Stage
0/I objects.  Since MIR-54 powers such a large outflow, it is probably
not a Class 0 (or stage 0) protostar, but rather a deeply embedded,
somewhat more evolved Class I protostar.

\section{Discussion and conclusions}

We find clear evidence for triggered star formation in the cometary
globule Ori\,I-2, with four young stars embedded in the front side of
the globule. The PMS object furthest away from the bright rim and approximately
at the center of the globule,  MIR-54, is a young Class I or stage I
protostar.  It is most likely responsible for the large parsec scale
Herbig-Haro outflow, with its counter flow seen as a bright H$\alpha$
outflow breaking out on the western side of the globule. However,
MIR-52, the heavily obscured H$\alpha$ emission line star $\sim$
20\arcsec\ south of MIR-54, also appears to drive an outflow, which is
approximately aligned with the outflow from MIR-54.  To the east both
outflows overlap, and either star could therefore be the exciting star
for the large HH outflow. MIR-89, which is located in the bright rim, is
a more evolved Class II object, which  most likely excites the HH
object, HH\,992, which we discovered in this study. HH\,992 lies outside
the globule, $\sim$ 65\arcsec\ east of the star and south of the large
Herbig-Haro outflow,  HH\,289.  HH\,992 breaks up into three knots, and
must be rather low excitation, since it is not seen in the H$\alpha$
image.  

\begin{figure*}[b]
\begin{center}
\resizebox{\hsize}{!}{\includegraphics[angle=0,width=8.0cm,angle=0]{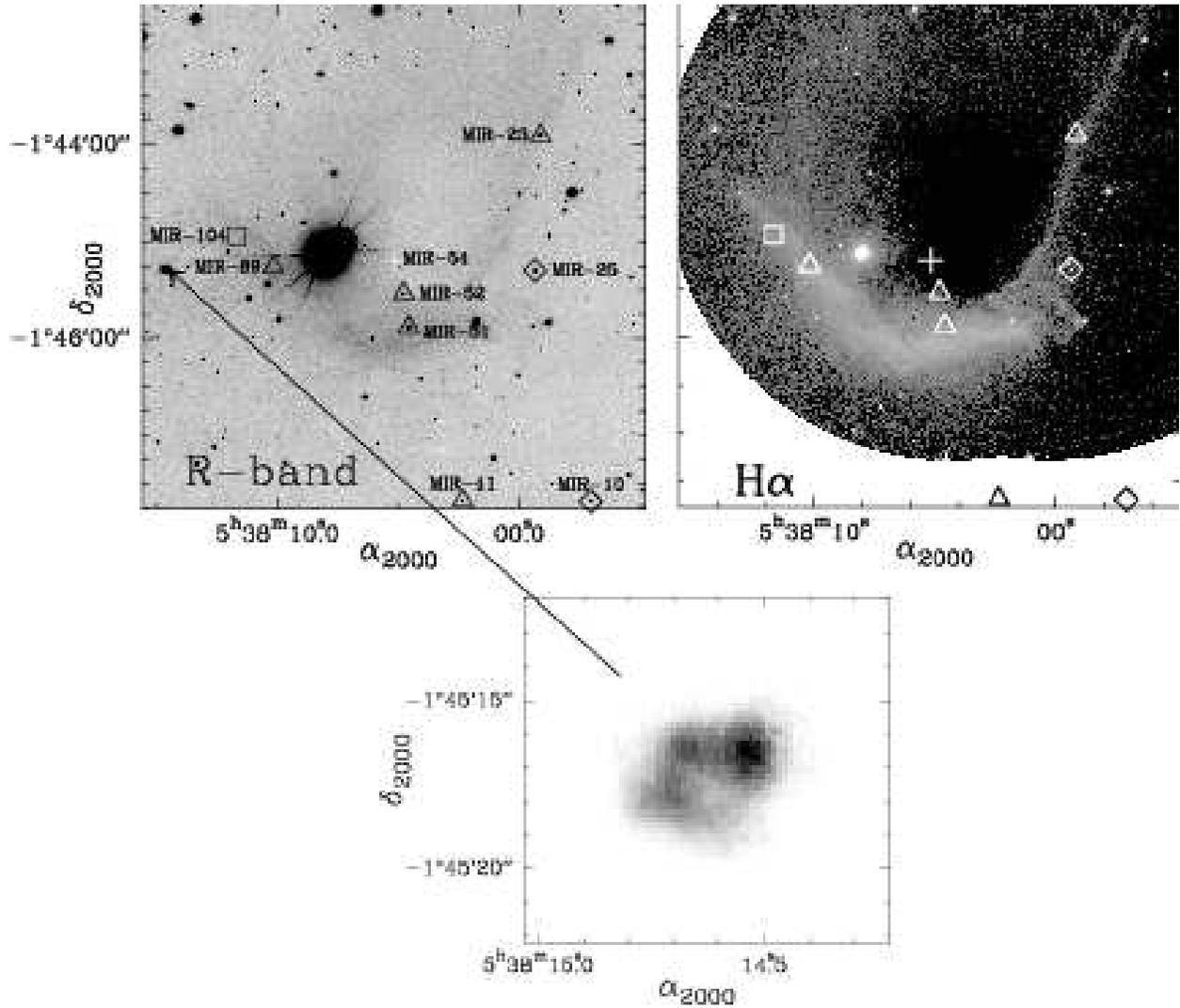}}
\caption{({\em Top left}) The deep $R$-band  and ({\em Top right}) the
H$\alpha$ emission images of Ori\,I-2 by NOT.  The crosses and
triangles indicate the sources securely identified as Class 0/I and
Class II objects respectively in Figure~\ref{fig_iracmipscol}, the
open diamonds show the candidate Class II objects and the open square
indicates the location of MIR-104.  The {\em bottom} panel shows a
closer view of the newly discovered HH object.
\label{fig_halpha_rband}}
\end{center}
\end{figure*}

Another candidate young star, MIR-104, which is close to the globule and
near MIR-89 (Figure~\ref{fig_halpha_rband}) may also have formed in the
globule.  However, MIR-104 was only detected at 5.8 and 8 \micron\ and
we therefore have insufficient information to determine whether it
really is a PMS object. If it is, it could  have formed inside the
globule before the gas was ablated away by the strong UV radiation from
$\sigma$ Ori.  

We identify two infrared excess stars, one of which is far to the south, $\sim$2\arcmin\  from the
globule and the other is seen in projection against the bright rim in
the tail region of the globule. Both these sources have zero or very
little foreground extinction and  are probably unrelated to the globule, although we cannot exclude the possibility that they could have formed in the
globule in an earlier epoch of star formation. We
also find two other candidate PMS stars in the field, which may have
weak mid-IR excesses. All these stars appear older, i.e. they are
pre-transitional or transitional disk stars, which now are part of
$\sigma$-Orionis cluster, which has an age of $\sim$ 3 Myr.

Models of cometary globules can successfully explain their morphology
and suggest that they form stars through Radiation Driven Implosion
(RDI)   \citep{lefloch1994,miao2006}.  This was already qualitatively
discussed by \citet{Reipurth83}, who showed that cometary globules are
active sites of star formation. In these models cometary globules are
denser cores in the molecular cloud surrounding an \ion{H}{2} region,
which get exposed to the UV radiation from the central OB association,
when the \ion{H}{2} region expands into it. The UV flux of the OB
association ionizes the external layers of the core facing the OB
association and an ionization front is formed at the front surface.  The
gas is heated during the ionization and the temperature increases. The
increased pressure due to the hot ionized gas drives an isothermal shock
into the cloud, which compresses the neutral gas in the core. At the
same time, the ionized and heated gas at the surface of the core flows
radially away from the surface and forms an  evaporating layer
surrounding the surface of the core. Due to the curvature of the front
surface of the core the shock waves preceding the ionization front
converge and collide at their focus, causing the interior of core to be
compressed to much higher densities. This causes the central part of the
core to become gravitationally unstable, triggering the formation of a
star.

The observations of \citet{ikeda2008} of the isolated cometary globule
BRC\,37 in the \ion{H}{2} region IC\,1396 appears to match this model
rather well. Their observations show a string of PMS objects along the
symmetry axis of the globule, with the oldest being in front of the
globule, and the youngest object still embedded in the head of the
globule, suggesting sequential formation of young stars as the
ionization front proceeds towards the tail of the globule. For some of
the  other cometary globules in the same \ion{H}{2} region, like
IC\,1396\,N (BRC\,38), this scenario is less clear. \citet{getman2007}
found an elongated clustering of X-ray sources aligned with the
globule, with the youngest stars still embedded inside it, consistent
with the RDI model for triggered star formation. However
\citet{beltran2009}, who studied the same cometary globule using deep
J, H, K$^\prime$, and narrowband H$_2$ 2.12~\micron\ imaging and
previously published mm imaging did not find any color or age gradient
in the south-north direction, i.e. in the direction of the globule.
They therefore conclude that not all star formation in this globule
can be explained in terms of triggering. We can think of two possible
explanations for why IC\,1396\,N does not appear to show the same type
of age progression and sequential triggering as seen in BRC\,37. If
the X-ray sources in front of the globule are late Class II objects,
i.e. transitional disk like objects, or Class III objects, they will
show no color excess in the near-IR and therefore appear as field
stars. Secondly, IC\,1396\,N is not a well defined isolated cometary
globule with a simple head/tail structure, suggesting that it could
consist of more than one cometary globule. 

Ori\,I-2, the cometary globule studied in this paper, does appear more
similar to BRC\,38. It shows clear evidence for triggered star
formation, with the youngest object, MIR-54, being furthest away from the
front side of the globule. The two other young Class II objects are
definitely older and closer to the bright rim and the ionizing star,
$\sigma$ Orionis. All  three stars are also close to the symmetry
axis of the globule, as predicted by the RDI model. MIR-89 and
possibly MIR-104, which both lie on the eastern side of the globule,
do not appear to have formed by  focussed compression along the
symmetry axis of the globule. Other processes, like Rayleigh-Taylor
instabilities in the bright rim, may also play a role and form
low-mass stars like MIR-89. A similar situation is seen in the
Horsehead, a bright rimmed elephant trunk structure east of $\sigma$
Orionis, which also shows some evidence for sequential star formation
triggered by the expanding \ion{H}{2} region \citep{mookerjea2009}.
These young PMS stars are also low-mass stars, which appear to have
formed in the bright rim by Rayleigh-Taylor instabilities,
since hardly any of them lie close to the symmetry axis of the Horsehead. 

Several, if not all of the young stars in Ori\,I-2 drive outflows. It
is interesting to note that all of the outflows are roughly
perpendicular to the symmetry axis of the globule. It maybe a
coincidence, but other cometary globules with embedded protostars,
like BRC\,37, show a similar behavior. If this is generally true, it
would provide additional support to the focussed RDI model, which
predicts that the gas inside a cometary globule will be compressed in
a ridge along the symmetry axis of the globule. When density
condensations, i.e. cores in this ridge become gravitationally
unstable and collapse, the collapse would preferentially be towards
the symmetry axis, therefore aligning the protoplanetary disk along
the symmetry axis. Since outflows are driven by disks, one would
therefore see outflows which are perpendicular to the symmetry axis of
the cometary globule.

Ori\,I-2 is thus a low mass star forming cometary globule. The star
formation in this region appears to be triggered by a combination of
RDI and Rayleigh-Taylor instabilities in the rim of the globule. In
the deep R-band images for the first time we detect HH~992, which
based on its non-detection in the \halpha\ image is most likely an
extremely low excitation HH object.  Follow-up optical and infrared
spectroscopic observations to understand the true nature of HH~992 as
well as MIR-52, the sole \halpha\ emission star in the field, are
needed.  We find that many of the young stars drive outflows which are
aligned perpendicular to the symmetry axis of the globule. Higher
spatial resolution imaging of the outflows in Ori\,I-2 in
near-infrared and millimeter would provide structural details of the
outflows and lead to correct identification of the sources exciting
these outflows. 

\acknowledgements
Based on observations made with the Nordic Optical Telescope, operated
on the island of La Palma jointly by Denmark, Finland, Iceland,
Norway, and Sweden, in the Spanish Observatorio del Roque de los
Muchachos of the Instituto de Astrofisica de Canarias.  We thank Bo Reipurth
for constructive criticism, which helped us improve the paper.

\newpage

\setcounter{table}{0}
\begin{deluxetable}{lccrcccccccc}
\tabletypesize{\scriptsize}
\rotate
\tablecaption{Mid-infrared sources in Ori I-2
\label{tab_mirsrc}}
\tablewidth{0pt}
\tablehead{
\colhead{Source} &
\colhead{$\alpha_{2000}$}&
\colhead{$\delta_{2000}$}&Optical&&
\colhead{2MASS}&&&&
\colhead{IRAC}&&
\colhead{MIPS}\\
& & & $R$\phantom{R} & $J$ & $H$ & $K_{\rm s}$ & $F_{3.6}$ & $F_{4.5}$ & $F_{5.8}$ &
$F_{8.0}$ & $F_{24}$\\
Ori\,I-2&&&mag&mag&mag&mag&mJy&mJy&mJy&mJy&mJy\\}
\startdata
MIR-1  &  5:37:55.30 &  -1:43:39.2 &  17.5 &   14.71$\pm$0.03  &    14.00$\pm$ 0.03  &    13.82$\pm$ 0.06 &     1.08$\pm$0.01&     0.71 $\pm$ 0.01&     0.48 $\pm$ 0.01&     0.32 $\pm$ 0.02&  \ldots\\
MIR-2  &  5:37:55.41 &  -1:45:22.1 & 20.3 &  \ldots  & \ldots  & \ldots &     0.18$\pm$0.01&     0.13 $\pm$ 0.00& \ldots& \ldots&  \ldots\\
MIR-3  &  5:37:55.43 &  -1:43:16.4 & 16.3 &     14.66$\pm$0.04  &    14.15$\pm$ 0.04  &    14.14$\pm$ 0.06 &     0.72$\pm$0.01&     0.44 $\pm$ 0.01&     0.30 $\pm$ 0.01&     0.17 $\pm$ 0.02&  \ldots\\
MIR-4  &  5:37:55.47 &  -1:47:24.7 & 17.2 &  14.84$\pm$0.03  &    14.05$\pm$ 0.04  &    13.85$\pm$ 0.05 &     0.95$\pm$0.02&     0.59 $\pm$ 0.01&     0.43 $\pm$ 0.03& \ldots&  \ldots\\
MIR-5  &  5:37:55.55 &  -1:46:04.4 & 22.1 & \ldots  & \ldots  & \ldots &     0.08$\pm$0.01&     0.07 $\pm$ 0.00& \ldots& \ldots&  \ldots\\
MIR-6  &  5:37:55.93 &  -1:45:56.4 & 21.2 &  \ldots  & \ldots  & \ldots &     0.07$\pm$0.00&     0.05 $\pm$ 0.00& \ldots& \ldots&  \ldots\\
MIR-7  &  5:37:55.94 &  -1:42:33.4 & 17.3 &     14.78$\pm$0.03  &    14.13$\pm$ 0.04  &    13.85$\pm$ 0.05 & \nodata\tablenotemark{a}&     0.62 $\pm$ 0.01&  \nodata\tablenotemark{a}& \ldots&  \ldots\\
MIR-8  &  5:37:56.30 &  -1:42:55.4 & 15.2 &      12.52$\pm$0.03  &    11.81$\pm$ 0.02  &    11.63$\pm$ 0.02 &     7.65$\pm$0.16&     4.81 $\pm$ 0.02&     3.03 $\pm$ 0.03&     1.78 $\pm$ 0.02&  \ldots\\
MIR-9  &  5:37:56.39 &  -1:45:15.2 & 23.3 & \ldots  & \ldots  & \ldots &     0.09$\pm$0.00&     0.06 $\pm$ 0.00& \ldots& \ldots&  \ldots\\
MIR-10 &  5:37:57.04 &  -1:47:40.9 & 18.2  &   16.72$\pm$0.13  &    16.18$\pm$ 0.15  &    16.81$\pm$\ldots &     0.13$\pm$0.00&     0.08 $\pm$ 0.00&     0.10 $\pm$ 0.01& \ldots&  \ldots\\
MIR-11 &  5:37:57.28 &  -1:44:10.5 & \ldots & \ldots  & \ldots  & \ldots & \ldots& \ldots& \ldots&     0.17 $\pm$ 0.02&  \ldots\\
MIR-12 &  5:37:57.80 &  -1:44:29.5 &  14.0 &   12.69$\pm$0.03  &    12.38$\pm$ 0.02  &    12.37$\pm$ 0.03 &     3.66$\pm$0.02&     2.32 $\pm$ 0.01&     1.49 $\pm$ 0.02&     0.86 $\pm$ 0.02&  \ldots\\
MIR-13 &  5:37:57.99 &  -1:47:27.4 &\ldots & \ldots  & \ldots  & \ldots &     0.05$\pm$0.00&     0.04 $\pm$ 0.00& \ldots& \ldots&  \ldots\\
MIR-14 &  5:37:58.01 &  -1:45:23.4 & 22.2& \ldots  & \ldots  & \ldots &     0.08$\pm$0.00&     0.05 $\pm$ 0.00& \ldots& \ldots&  \ldots\\
MIR-15 &  5:37:58.12 &  -1:44:40.7 &   17.9 &   16.63$\pm$0.12  &    16.27$\pm$ 0.16  &    14.96$\pm$\ldots &     0.12$\pm$0.00&     0.07 $\pm$ 0.00& \ldots& \ldots&  \ldots\\
MIR-16 &  5:37:58.29 &  -1:42:37.4 &19.3 & \ldots  & \ldots  & \ldots &  \nodata\tablenotemark{a}&     0.13 $\pm$ 0.00&  \nodata\tablenotemark{a}&     0.29 $\pm$ 0.02&  \ldots\\
MIR-17 &  5:37:58.38 &  -1:47:30.3 & 20.0 &  16.73$\pm$0.13  &    16.31$\pm$ 0.16  &    15.50$\pm$ 0.22 &     0.18$\pm$0.00&     0.12 $\pm$ 0.00&     0.14 $\pm$ 0.01& \ldots&  \ldots\\
MIR-18 &  5:37:58.64 &  -1:47:30.6 &21.5 & \ldots  & \ldots  & \ldots &     0.15$\pm$0.00&     0.10 $\pm$ 0.00& \ldots& \ldots&  \ldots\\
MIR-19 &  5:37:58.77 &  -1:44:56.3 & 18.1 &     16.81$\pm$0.14  &    16.34$\pm$ 0.17  &    15.95$\pm$\ldots &     0.11$\pm$0.00&     0.07 $\pm$ 0.00& \ldots& \ldots&  \ldots\\
MIR-20 &  5:37:58.78 &  -1:45:50.0 & 15.6 &     13.10$\pm$0.03  &    12.40$\pm$ 0.02  &    12.19$\pm$ 0.02 &     4.32$\pm$0.02&     2.70 $\pm$ 0.01&     1.95 $\pm$ 0.02&     1.06 $\pm$ 0.02&  \ldots\\
MIR-21 &  5:37:58.86 &  -1:47:28.1 & 23.1 & \ldots  & \ldots  & \ldots &     0.11$\pm$0.00&     0.07 $\pm$ 0.00& \ldots& \ldots&  \ldots\\
MIR-22 &  5:37:59.07 &  -1:44:14.2 & \ldots& \ldots  & \ldots  & \ldots & \ldots& \ldots& \ldots&     0.14 $\pm$ 0.02&  \ldots\\
MIR-23 &  5:37:59.13 &  -1:43:54.2 & 16.9 &     14.53$\pm$0.03  &    13.77$\pm$ 0.04  &    13.63$\pm$ 0.05 &     1.26$\pm$0.01&     1.16 $\pm$ 0.10&     1.04 $\pm$ 0.02&     1.50 $\pm$ 0.02&  \ldots\\
MIR-24 &  5:37:59.24 &  -1:46:35.5 & \ldots& \ldots  & \ldots  & \ldots &     0.04$\pm$0.00&     0.03 $\pm$ 0.00& \ldots& \ldots&  \ldots\\
MIR-25 &  5:37:59.35 &  -1:45:17.9 & 16.9 &     15.26$\pm$0.04  &    14.70$\pm$ 0.04  &    14.57$\pm$ 0.07 &     0.40$\pm$0.01&     0.26 $\pm$ 0.01&     0.16 $\pm$ 0.01&     0.16 $\pm$ 0.02&  \ldots\\
MIR-26 &  5:37:59.36 &  -1:44:32.1 & 20.5 &    17.04$\pm$0.16  &    16.44$\pm$ 0.18  &    15.54$\pm$ 0.22 &     0.16$\pm$0.00&     0.11 $\pm$ 0.00&     0.09 $\pm$ 0.01& \ldots&  \ldots\\
MIR-27 &  5:37:59.49 &  -1:45:52.2 & 19.0&    16.15$\pm$0.07  &    15.57$\pm$ 0.10  &    15.25$\pm$ 0.16 &     0.23$\pm$0.00&     0.15 $\pm$ 0.00&     0.10 $\pm$ 0.01& \ldots&  \ldots\\
MIR-28 &  5:37:59.53 &  -1:46:49.2 & \ldots& \ldots  & \ldots  & \ldots &     0.12$\pm$0.00&     0.08 $\pm$ 0.00& \ldots& \ldots&  \ldots\\
MIR-29 &  5:38:00.00 &  -1:44:30.6 & 18.3  &     15.98$\pm$0.07  &    15.24$\pm$ 0.07  &    15.23$\pm$ 0.17 &     0.35$\pm$0.00&     0.19 $\pm$ 0.00& \ldots& \ldots&  \ldots\\
MIR-30 &  5:38:00.04 &  -1:46:22.5 &  18.6 & \ldots  & \ldots  & \ldots &     0.05$\pm$0.00&     0.06 $\pm$ 0.00& \ldots& \ldots&  \ldots\\
MIR-31 &  5:38:00.44 &  -1:42:59.4 &  19.5 & \ldots  & \ldots  & \ldots &     0.06$\pm$0.00&     0.05 $\pm$ 0.00& \ldots& \ldots&  \ldots\\
MIR-32 &  5:38:00.90 &  -1:46:44.9 &  \ldots & \ldots  & \ldots  & \ldots &     0.07$\pm$0.00&     0.04 $\pm$ 0.00& \ldots& \ldots&  \ldots\\
MIR-33 &  5:38:01.03 &  -1:44:33.7 &  \ldots & \ldots  & \ldots  & \ldots &     0.15$\pm$0.00&     0.11 $\pm$ 0.00& \ldots& \ldots&  \ldots\\
MIR-34 &  5:38:01.06 &  -1:47:14.1 &  16.4 &   14.91$\pm$0.04  &    14.30$\pm$ 0.04  &    14.46$\pm$ 0.07 &     0.51$\pm$0.00&     0.31 $\pm$ 0.00&     0.24 $\pm$ 0.01&     0.15 $\pm$ 0.02&  \ldots\\
MIR-35 &  5:38:01.07 &  -1:44:50.8 & 22.5 &  \ldots  & \ldots  & \ldots &     0.09$\pm$0.00&     0.07 $\pm$ 0.00& \ldots& \ldots&  \ldots\\
MIR-36 &  5:38:01.07 &  -1:43:51.1 & \ldots &  \ldots  & \ldots  & \ldots &     0.05$\pm$0.00&     0.05 $\pm$ 0.00& \ldots& \ldots&  \ldots\\
MIR-37 &  5:38:01.60 &  -1:44:32.5 & \ldots  & \ldots  & \ldots  & \ldots &     0.06$\pm$0.00&     0.04 $\pm$ 0.00& \ldots& \ldots&  \ldots\\
MIR-38 &  5:38:01.83 &  -1:45:50.1 &  15.1  &   12.45$\pm$0.03  &    11.82$\pm$ 0.02  &    11.62$\pm$ 0.03 &     7.88$\pm$0.03&     4.40 $\pm$ 0.02&     4.07 $\pm$ 0.03&     3.86 $\pm$ 0.03&  \ldots\\
MIR-39 &  5:38:02.10 &  -1:44:10.9 &  23.2 &  \ldots  & \ldots  & \ldots &     0.06$\pm$0.00&     0.04 $\pm$ 0.00& \ldots& \ldots&  \ldots\\
MIR-40 &  5:38:02.24 &  -1:47:30.1 & \ldots & \ldots  & \ldots  & \ldots &     0.05$\pm$0.00&     0.05 $\pm$ 0.00& \ldots& \ldots&  \ldots\\
MIR-41 &  5:38:02.33 &  -1:47:40.4 & 19.1  &    14.09$\pm$0.03  &    13.49$\pm$ 0.03  &    13.10$\pm$ 0.03 &     3.48$\pm$0.11&     3.21 $\pm$ 0.30&     2.88 $\pm$ 0.03&     2.55 $\pm$ 0.06&      2.40 $\pm$ 0.05\\
MIR-42 &  5:38:02.35 &  -1:43:47.3 & \ldots & \ldots  & \ldots  & \ldots &     0.08$\pm$0.00&     0.05 $\pm$ 0.00& \ldots& \ldots&  \ldots\\
MIR-43 &  5:38:02.36 &  -1:44:01.8 & 22.7& \ldots & \ldots  & \ldots &     0.02$\pm$0.00&     0.03 $\pm$ 0.00& \ldots& \ldots&  \ldots\\
MIR-44 &  5:38:02.87 &  -1:42:36.7 & 17.4 &    13.12$\pm$0.03  &    12.58$\pm$ 0.02  &    12.23$\pm$ 0.03 &  \nodata\tablenotemark{b}&     3.34 $\pm$ 0.02&     2.13 $\pm$ 0.04&     1.26 $\pm$ 0.03&  \ldots\\
MIR-45 &  5:38:03.42 &  -1:45:15.1 & \ldots & \ldots  & \ldots  & \ldots &     0.09$\pm$0.00&     0.04 $\pm$ 0.00& \ldots& \ldots&  \ldots\\
MIR-46 &  5:38:03.57 &  -1:44:15.6 & \ldots& \ldots  & \ldots  & \ldots &     0.08$\pm$0.00&     0.06 $\pm$ 0.00& \ldots& \ldots&  \ldots\\
MIR-47 &  5:38:03.87 &  -1:45:57.5 & 17.6 &    16.01$\pm$0.07  &    15.40$\pm$ 0.08  &    15.40$\pm$ 0.17 &     0.28$\pm$0.00&     0.16 $\pm$ 0.00& \ldots& \ldots&  \ldots\\
MIR-48 &  5:38:04.10 &  -1:46:25.1 &18.5 &  \ldots  & \ldots  & \ldots &     0.09$\pm$0.00&     0.05 $\pm$ 0.00& \ldots& \ldots&  \ldots\\
MIR-49 &  5:38:04.33 &  -1:44:37.9 & \ldots &   17.27$\pm$\ldots  &    16.68$\pm$ 0.26  &    15.20$\pm$ 0.16 &     0.63$\pm$0.01&     0.54 $\pm$ 0.01&     0.43 $\pm$ 0.01& \ldots&  \ldots\\
MIR-50 &  5:38:04.43 &  -1:45:50.3 &\ldots & \ldots  & \ldots  & \ldots &     0.15$\pm$0.00&     0.09 $\pm$ 0.00&     0.76 $\pm$ 0.01& \ldots&  \ldots\\
MIR-51 &  5:38:04.56 &  -1:45:53.0 & 17.3  &   15.26$\pm$0.04  &    14.72$\pm$ 0.04  &    14.64$\pm$ 0.10 &     0.66$\pm$0.01&     0.76 $\pm$ 0.10&     1.67 $\pm$ 0.37&     2.72 $\pm$ 0.03&  \ldots\\
MIR-52 &  5:38:04.78 &  -1:45:32.2 & 19.9 &  13.56$\pm$0.04  &    12.73$\pm$ 0.04  &    12.16$\pm$ 0.03 &     7.18$\pm$0.03&     5.13 $\pm$ 0.12&     4.00 $\pm$ 0.38&     2.91 $\pm$ 0.03&     11.37 $\pm$ 0.05\\
MIR-53 &  5:38:05.10 &  -1:45:41.9 & 22.8 & \ldots   & \ldots  & \ldots &     0.09$\pm$0.00&     0.02 $\pm$ 0.00& \ldots& \ldots&  \ldots\\
MIR-54 &  5:38:05.15 &  -1:45:12.4 & \ldots &  \ldots  & \ldots  & \ldots &     3.10$\pm$0.02&     4.97 $\pm$ 0.04&    15.73 $\pm$ 0.06&    13.63 $\pm$ 0.06&    231.90 $\pm$ 0.44\\
MIR-55 &  5:38:05.19 &  -1:46:18.9 & 20.7&  \ldots  & \ldots  & \ldots &     0.08$\pm$0.00&     0.06 $\pm$ 0.00& \ldots& \ldots&  \ldots\\
MIR-56 &  5:38:05.40 &  -1:47:05.8 &22.5 &  \ldots  & \ldots  & \ldots &     0.10$\pm$0.00&     0.06 $\pm$ 0.00& \ldots& \ldots&  \ldots\\
MIR-57 &  5:38:05.50 &  -1:42:43.1 & 21.7 &  \ldots  & \ldots  & \ldots & \ldots&     0.10 $\pm$ 0.00& \ldots& \ldots&  \ldots\\
MIR-58 &  5:38:05.83 &  -1:42:36.3 & 14.1 &    12.45$\pm$0.03  &    11.94$\pm$ 0.02  &    11.84$\pm$ 0.03 &     5.86$\pm$0.28&     3.47 $\pm$ 0.02&     2.26 $\pm$ 0.03&     1.28 $\pm$ 0.02&  \ldots\\
MIR-59 &  5:38:05.86 &  -1:47:04.1 & 19.1  &    16.95$\pm$0.15  &    16.26$\pm$ 0.16  &    15.70$\pm$ 0.23 &     0.13$\pm$0.00&     0.08 $\pm$ 0.00& \ldots& \ldots&  \ldots\\
MIR-60 &  5:38:05.90 &  -1:43:05.7 & 19.8 & \ldots  & \ldots  & \ldots &     0.07$\pm$0.00&     0.05 $\pm$ 0.00& \ldots& \ldots&  \ldots\\
MIR-61 &  5:38:06.06 &  -1:46:00.2 &18.2  &   15.93$\pm$0.07  &    15.34$\pm$ 0.07  &    15.06$\pm$ 0.13 &     0.39$\pm$0.00&     0.20 $\pm$ 0.00& \ldots& \ldots&  \ldots\\
MIR-62 &  5:38:06.26 &  -1:43:59.6 &\ldots & \ldots  & \ldots  & \ldots &     0.06$\pm$0.00&     0.04 $\pm$ 0.00& \ldots& \ldots&  \ldots\\
MIR-63 &  5:38:06.38 &  -1:45:02.1 &20.2 &   16.71$\pm$0.13  &    16.23$\pm$ 0.15  &    15.40$\pm$ 0.18 &     0.29$\pm$0.00&     0.02 $\pm$ 0.00& \ldots& \ldots&  \ldots\\
MIR-64 &  5:38:06.38 &  -1:46:36.4 &16.6 &    13.97$\pm$0.03  &    13.24$\pm$ 0.02  &    12.99$\pm$ 0.03 &     2.06$\pm$0.01&     1.30 $\pm$ 0.01&     0.93 $\pm$ 0.02&     0.47 $\pm$ 0.02&  \ldots\\
MIR-65 &  5:38:06.65 &  -1:46:55.5 & 20.8 & \ldots  & \ldots  & \ldots &     0.10$\pm$0.00&     0.07 $\pm$ 0.00& \ldots& \ldots&  \ldots\\
MIR-66 &  5:38:06.77 &  -1:47:14.8 & 21.6 & \ldots  & \ldots  & \ldots &     0.06$\pm$0.00&     0.05 $\pm$ 0.00& \ldots& \ldots&  \ldots\\
MIR-67 &  5:38:06.84 &  -1:47:03.8 &\ldots & \ldots  & \ldots  & \ldots &     0.04$\pm$0.00&     0.03 $\pm$ 0.00& \ldots& \ldots&  \ldots\\
MIR-68 &  5:38:06.99 &  -1:43:04.0 &20.1 &    16.89$\pm$0.14  &    16.07$\pm$ 0.14  &    15.77$\pm$ 0.26 &     0.14$\pm$0.00&     0.10 $\pm$ 0.00& \ldots& \ldots&  \ldots\\
MIR-69 &  5:38:07.25 &  -1:45:59.7 & 20.7 & \ldots  & \ldots  & \ldots &     0.13$\pm$0.00&     0.06 $\pm$ 0.00& \ldots& \ldots&  \ldots\\
MIR-70 &  5:38:07.27 &  -1:47:13.0 & 23.3 & \ldots  & \ldots  & \ldots &     0.05$\pm$0.00&     0.04 $\pm$ 0.00& \ldots& \ldots&  \ldots\\
MIR-71 &  5:38:07.64 &  -1:42:46.8 & 18.5  &   16.94$\pm$0.15  &    16.15$\pm$ 0.14  &    15.86$\pm$\ldots &     0.12$\pm$0.00&     0.07 $\pm$ 0.00& \ldots& \ldots&  \ldots\\
MIR-72 &  5:38:07.65 &  -1:44:17.9 & 16.2  &   13.69$\pm$0.03  &    13.01$\pm$ 0.03  &    12.76$\pm$ 0.04 &     2.39$\pm$0.02&     1.78 $\pm$ 0.01&     1.25 $\pm$ 0.02&     0.65 $\pm$ 0.02&  \ldots\\
MIR-73 &  5:38:08.01 &  -1:45:07.7 &  S &    8.48$\pm$0.03  &     8.55$\pm$ 0.03  &     8.57$\pm$ 0.02 &    99.1$\pm$0.2\phantom{0}&    59.6 $\pm$ 0.2\phantom{0}&    41.4 $\pm$ 0.1\phantom{0}&    21.8 $\pm$ 0.1\phantom{0}&      6.32 $\pm$ 0.05\\
MIR-74 &  5:38:08.11 &  -1:42:43.5 & 20.1 &    16.94$\pm$0.15  &    16.20$\pm$ 0.15  &    15.47$\pm$ 0.19 &     0.16$\pm$0.01&     0.11 $\pm$ 0.00&     0.04 $\pm$ 0.01& \ldots&  \ldots\\
MIR-75 &  5:38:08.19 &  -1:43:16.0 & 16.3 &    14.77$\pm$0.03  &    14.27$\pm$ 0.04  &    14.08$\pm$ 0.06 &     0.64$\pm$0.01&     0.45 $\pm$ 0.01&    0.30 $\pm$ 0.01&     0.12 $\pm$ 0.02&  \ldots\\
MIR-76 &  5:38:08.68 &  -1:43:01.1 & 18.2  &    15.94$\pm$0.06  &    15.16$\pm$ 0.06  &    14.80$\pm$ 0.10 &     0.34$\pm$0.00&     0.22 $\pm$ 0.00&     0.08 $\pm$ 0.01& \ldots&  \ldots\\
MIR-77 &  5:38:08.88 &  -1:43:28.3 & 22.6 &  \ldots  & \ldots  & \ldots &     0.03$\pm$0.00&     0.03 $\pm$ 0.00& \ldots& \ldots&  \ldots\\
MIR-78 &  5:38:08.92 &  -1:44:48.2 & 18.2 &  16.61$\pm$0.11  &    16.11$\pm$ 0.14  &    16.81$\pm$\ldots &     0.11$\pm$0.00&     0.07 $\pm$ 0.00& \ldots& \ldots&  \ldots\\
MIR-79 &  5:38:09.14 &  -1:45:06.7 & \ldots & \ldots  & \ldots  & \ldots &     0.06$\pm$0.00&     0.05 $\pm$ 0.00& \ldots& \ldots&  \ldots\\
MIR-80 &  5:38:09.40 &  -1:44:20.7 &\ldots & \ldots  & \ldots  & \ldots &     0.07$\pm$0.00&     0.13 $\pm$ 0.00& \ldots& \ldots&  \ldots\\
MIR-81 &  5:38:09.66 &  -1:47:16.8 &17.0 &    15.12$\pm$0.04  &    14.64$\pm$ 0.04  &    14.41$\pm$ 0.08 &     0.51$\pm$0.00&     0.33 $\pm$ 0.00&     0.27 $\pm$ 0.01&     0.14 $\pm$ 0.02&  \ldots\\
MIR-82 &  5:38:09.69 &  -1:44:51.0 & \ldots & \ldots  & \ldots  & \ldots &     0.03$\pm$0.00&     0.04 $\pm$ 0.00& \ldots& \ldots&  \ldots\\
MIR-83 &  5:38:09.78 &  -1:44:42.6 & 19.4 & \ldots  & \ldots  & \ldots &     0.06$\pm$0.00&     0.05 $\pm$ 0.00& \ldots& \ldots&  \ldots\\
MIR-84 &  5:38:09.93 &  -1:45:47.5 &16.2 &     14.58$\pm$0.03  &    14.19$\pm$ 0.03  &    14.12$\pm$ 0.06 &     0.67$\pm$0.01&     0.44 $\pm$ 0.01&     0.30 $\pm$ 0.01&     0.21 $\pm$ 0.02&  \ldots\\
MIR-85 &  5:38:09.96 &  -1:45:53.0 &\ldots & \ldots  & \ldots  & \ldots &     0.02$\pm$0.00&     0.02 $\pm$ 0.00& \ldots& \ldots&  \ldots\\
MIR-86 &  5:38:09.97 &  -1:47:00.6 & 20.3 & \ldots  & \ldots  & \ldots &     0.11$\pm$0.00&     0.07 $\pm$ 0.00& \ldots& \ldots&  \ldots\\
MIR-87 &  5:38:10.01 &  -1:43:35.6 &17.8  &   16.23$\pm$0.09  &    15.94$\pm$ 0.13  &    15.75$\pm$ 0.25 &     0.16$\pm$0.00&     0.10 $\pm$ 0.00& \ldots& \ldots&  \ldots\\
MIR-88 &  5:38:10.02 &  -1:45:38.1 & 23.5 & \ldots  & \ldots  & \ldots &     0.70$\pm$0.09&     0.35 $\pm$ 0.10&     0.12 $\pm$ 0.01&     0.15 $\pm$ 0.02&  \ldots\\
MIR-89 &  5:38:10.16 &  -1:45:15.6 & \ldots & \ldots  & \ldots  & \ldots &     0.14$\pm$0.00&     0.58 $\pm$ 0.10&     0.73 $\pm$ 0.01&     2.91 $\pm$ 0.02&     21.40 $\pm$ 0.05\\
MIR-90 &  5:38:10.30 &  -1:44:15.4 & \ldots& \ldots  & \ldots  & \ldots &     0.05$\pm$0.00&     0.06 $\pm$ 0.00& \ldots& \ldots&  \ldots\\
MIR-91 &  5:38:10.31 &  -1:43:44.8 & \ldots& \ldots  & \ldots  & \ldots &     0.11$\pm$0.00&     0.07 $\pm$ 0.00& \ldots& \ldots&  \ldots\\
MIR-92 &  5:38:10.35 &  -1:45:59.4 & \ldots& \ldots  & \ldots  & \ldots &     0.01$\pm$0.00&     0.02 $\pm$ 0.00& \ldots& \ldots&  \ldots\\
MIR-93 &  5:38:10.41 &  -1:45:26.4 & 19.2 &   16.02$\pm$0.09  &    15.13$\pm$ 0.08  &    14.62$\pm$ 0.11 &     0.63$\pm$0.01&     0.35 $\pm$ 0.00&     0.29 $\pm$ 0.01&     0.51 $\pm$ 0.02&  \ldots\\
MIR-94 &  5:38:10.52 &  -1:43:13.7 & 18.6 &   17.11$\pm$0.18  &    16.03$\pm$ 0.14  &    15.54$\pm$\ldots &     0.12$\pm$0.00&     0.08 $\pm$ 0.00& \ldots& \ldots&  \ldots\\
MIR-95 &  5:38:10.69 &  -1:44:07.4 & \ldots& \ldots  & \ldots  & \ldots &     0.10$\pm$0.00&     0.06 $\pm$ 0.00& \ldots& \ldots&  \ldots\\
MIR-96 &  5:38:10.70 &  -1:45:03.9 & 23.7 & \ldots  & \ldots  & \ldots &     0.08$\pm$0.00&     0.05 $\pm$ 0.00& \ldots& \ldots&  \ldots\\
MIR-97 &  5:38:10.79 &  -1:44:11.6 & 22.9 & \ldots  & \ldots  & \ldots &     0.09$\pm$0.00&     0.06 $\pm$ 0.00& \ldots& \ldots&  \ldots\\
MIR-98 &  5:38:10.80 &  -1:44:44.5 & 19.2 & \ldots  & \ldots  & \ldots &     0.07$\pm$0.00&     0.04 $\pm$ 0.00& \ldots& \ldots&  \ldots\\
MIR-99 &  5:38:10.85 &  -1:46:57.6 & \ldots& \ldots  & \ldots  & \ldots &     0.09$\pm$0.00&     0.07 $\pm$ 0.00& \ldots& \ldots&  \ldots\\
MIR-100 &  5:38:10.85 &  -1:46:52.1 & \ldots & \ldots  & \ldots  & \ldots &     0.10$\pm$0.00&     0.05 $\pm$ 0.00& \ldots& \ldots&  \ldots\\
MIR-101 &  5:38:10.98 &  -1:44:03.1 &\ldots & \ldots  & \ldots  & \ldots &     0.07$\pm$0.00&     0.06 $\pm$ 0.00& \ldots& \ldots&  \ldots\\
MIR-102 &  5:38:11.08 &  -1:46:09.5 &\ldots & \ldots  & \ldots  & \ldots &     0.03$\pm$0.00&     0.04 $\pm$ 0.00& \ldots& \ldots&  \ldots\\
MIR-103 &  5:38:11.21 &  -1:45:35.0 &  16.9&   15.54$\pm$0.06  &    15.39$\pm$ 0.08  &    15.15$\pm$ 0.15 &     0.27$\pm$0.00&     0.17 $\pm$ 0.00&     0.07 $\pm$ 0.01& \ldots&  \ldots\\
MIR-104 &  5:38:11.66 &  -1:44:56.5 &\ldots & \ldots  & \ldots  & \ldots & \ldots& \ldots&     0.10 $\pm$ 0.01&     0.30 $\pm$ 0.02&  \ldots\\
MIR-105 &  5:38:11.71 &  -1:46:57.4 &\ldots & \ldots  & \ldots  & \ldots &     0.07$\pm$0.00&     0.05 $\pm$ 0.00& \ldots& \ldots&  \ldots\\
MIR-106 &  5:38:11.71 &  -1:44:48.1 & 23.1 & \ldots  & \ldots  & \ldots &     0.06$\pm$0.00&     0.04 $\pm$ 0.00& \ldots& \ldots&  \ldots\\
MIR-107 &  5:38:11.81 &  -1:43:01.1 & 17.6&    14.72$\pm$0.03  &    14.01$\pm$ 0.04  &    13.80$\pm$ 0.05 &     1.07$\pm$0.01&     0.74 $\pm$ 0.01&     0.45 $\pm$ 0.01&     0.25 $\pm$ 0.02&  \ldots\\
MIR-108 &  5:38:11.97 &  -1:46:38.8 &\ldots & \ldots  & \ldots  & \ldots &     0.17$\pm$0.00&     0.12 $\pm$ 0.00& \ldots& \ldots&  \ldots\\
MIR-109 &  5:38:11.99 &  -1:45:44.5 & 21.2 & \ldots  & \ldots  & \ldots &     0.05$\pm$0.00&     0.03 $\pm$ 0.00& \ldots& \ldots&  \ldots\\
MIR-110 &  5:38:12.01 &  -1:45:06.5 & 18.2 &   16.56$\pm$0.12  &    16.46$\pm$ 0.19  &    15.72$\pm$ 0.23 &     0.08$\pm$0.00&     0.05 $\pm$ 0.00& \ldots& \ldots&  \ldots\\
MIR-111 &  5:38:12.04 &  -1:43:36.6 & \ldots& \ldots  & \ldots  & \ldots &     0.06$\pm$0.00&     0.04 $\pm$ 0.00& \ldots& \ldots&  \ldots\\
MIR-112 &  5:38:12.16 &  -1:46:56.6 & 23.7 & \ldots  & \ldots  & \ldots &     0.06$\pm$0.00&     0.06 $\pm$ 0.00& \ldots& \ldots&  \ldots\\
MIR-113 &  5:38:12.44 &  -1:46:53.8 & \ldots & \ldots  & \ldots  & \ldots &     0.09$\pm$0.00&     0.08 $\pm$ 0.00& \ldots& \ldots&  \ldots\\
MIR-114 &  5:38:12.62 &  -1:47:07.5 & 17.5 &    15.60$\pm$0.05  &    15.29$\pm$ 0.08  &    14.95$\pm$ 0.12 &     0.32$\pm$0.00&     0.20 $\pm$ 0.00&     0.16 $\pm$ 0.01&     0.13 $\pm$ 0.02&  \ldots\\
MIR-115 &  5:38:13.32 &  -1:47:17.5 &  17.4 &   15.65$\pm$0.05  &    15.17$\pm$ 0.09  &    14.97$\pm$ 0.13 &     0.31$\pm$0.00&     0.21 $\pm$ 0.00&     0.14 $\pm$ 0.01& \ldots&  \ldots\\
MIR-116 &  5:38:13.37 &  -1:42:41.9 &  13.8 &   12.40$\pm$0.03  &    12.10$\pm$ 0.03  &    12.02$\pm$ 0.03 &     4.74$\pm$0.03&     3.14 $\pm$ 0.01&     1.93 $\pm$ 0.02&     1.10 $\pm$ 0.02&  \ldots\\
MIR-117 &  5:38:13.57 &  -1:43:36.1 & 16.4 &    15.15$\pm$0.04  &    14.67$\pm$ 0.06  &    14.98$\pm$ 0.15 &     0.44$\pm$0.01&     0.30 $\pm$ 0.00&     0.14 $\pm$ 0.01& \ldots&  \ldots\\
MIR-118 &  5:38:13.59 &  -1:44:27.8 & \ldots & \ldots  & \ldots  & \ldots &     0.08$\pm$0.00&     0.04 $\pm$ 0.00& \ldots& \ldots&  \ldots\\
MIR-119 &  5:38:13.69 &  -1:45:15.3 & 19.7 & \ldots  & \ldots  & \ldots &     0.12$\pm$0.00&     0.07 $\pm$ 0.00& \ldots& \ldots&  \ldots\\
MIR-120 &  5:38:14.08 &  -1:43:50.9 & 14.2 &    12.19$\pm$0.03  &    11.63$\pm$ 0.02  &    11.51$\pm$ 0.03 &     7.79$\pm$0.03&     4.90 $\pm$ 0.02&     3.21 $\pm$ 0.03&     1.84 $\pm$ 0.02&  \ldots\\
MIR-121 &  5:38:14.68 &  -1:44:53.1 & 23.5 & \ldots  & \ldots  & \ldots &     0.12$\pm$0.00&     0.07 $\pm$ 0.00& \ldots& \ldots&  \ldots\\
MIR-122 &  5:38:14.90 &  -1:46:07.9 &\ldots & \ldots  & \ldots  & \ldots &     0.06$\pm$0.00&     0.07 $\pm$ 0.00& \ldots& \ldots&  \ldots\\
MIR-123 &  5:38:14.92 &  -1:47:04.1 & 18.4 &    16.90$\pm$0.14  &    16.52$\pm$ 0.21  &    16.35$\pm$\ldots &     0.09$\pm$0.00&     0.06 $\pm$ 0.00& \ldots& \ldots&  \ldots\\
MIR-124 &  5:38:15.01 &  -1:45:58.6 & 20.4 & \ldots  & \ldots  & \ldots &     0.13$\pm$0.00&     0.10 $\pm$ 0.00& \ldots& \ldots&  \ldots\\
MIR-125 &  5:38:15.45 &  -1:45:56.1 & 20.4 & \ldots  & \ldots  & \ldots &     0.10$\pm$0.00&     0.09 $\pm$ 0.00& \ldots& \ldots&  \ldots\\
\enddata
\tablenotetext{a}{Outside the area mapped in 3.6 and 5.8~\micron{}}
\tablenotetext{b}{The star is at the edge of the  3.6~\micron-map and only
partially detected}\\

\end{deluxetable}

\begin{deluxetable}{lcc}
\tablecolumns{3}
\tablewidth{0pt}
\tablecaption{Coordinates of the three knots of HH\,992
\label{tab_hhobjects}}
\tablehead{
\colhead{Name} &  \colhead{$\alpha$(2000.0)} &  \colhead{$\delta$(2000.0)} 
}
\startdata
HH\,992A & 05:38:14.53 & -01:45:16.6  \\
HH\,992B & 05:38:14.67 & -01:45:16.3 \\
HH\,992C & 05:38:14.70 & -01:45:17.6 \\
\enddata
\end{deluxetable}
\begin{deluxetable}{crrrrrrrr}
\tabletypesize{\tiny}
\tablecolumns{9}
\tablewidth{0pt}
\tablecaption{Parameters derived from SED modeling using axisymmetric
radiation transfer models for the candidate PMS stars. All best fit models 
correspond to a distance of 400$\pm$50~pc. \av\ refers to the foreground 
extinction towards the source.
\label{tab_sedfits}}
\tablehead{
\colhead{Source} & \colhead{$T_{\ast}$} & \colhead{\mstar} & \colhead{$M_{\rm disk}$} & \colhead{$M_{\rm env}$} &
\colhead{$\dot{M}_{\rm env}$} & \colhead{$\dot{M}_{\rm disk}$} &
\colhead{$L$} & \colhead{\av}\\
\colhead{}& \colhead{K} & \colhead{\msun} & \colhead{\rm M$_{\odot}$} & \colhead{\rm M$_{\odot}$} & \colhead{\rm M$_{\odot}$~yr$^{-1}$} &\colhead{\rm M$_{\odot}$~yr$^{-1}$} &
\colhead{\rm L$_{\odot}$} &
}
\startdata
MIR-23 & 3110$\pm$18  & 0.13$\pm$0.01 &  1.7(-4)$\pm$5.3(-5) &
2.9(-6)$\pm$9.2(-7) & 0.0 & 1.9(-11)$\pm$6.2(-13) & 0.03$\pm$0.001 &
0.1$\pm$0.03\\
MIR-41 & 3750$\pm$184  & 0.56$\pm$0.12 &  5.0(-3)$\pm$3(-3) &
2.0(-6)$\pm$1.8(-6) & 0.0 & 1.4(-8)$\pm$7.1(-9) & 0.68$\pm$0.003
& 3.1$\pm$0.8\\
MIR-51 & 4500  & 1.56 &  2.0(-4) & 3.3(-4) & 1.9(-7) & 8.2(-10) & 6.5 &
1.9\\
MIR-52 & 2980  & 0.14 &  5.2(-5) & 9.7(-4) & 1.0(-7) & 3.1(-11) & 0.24 &
4.8\\
MIR-54 & 3380  & 0.30 &  2.5(-3) & 1.9(-1) & 5.9(-6) & 1.1(-8) &
1.36&12\\
MIR-89 & 3370 & 0.24 &  4.5(-3) & 2.1(-9) & 0.0 & 3.3(-8) &
0.3\phantom{0}& 50\\
\enddata
\end{deluxetable}

\newpage

\end{document}